\documentclass{aa}  

\usepackage{graphicx}
\usepackage[labelfont=bf,font=small,tableposition=top]{caption}
\usepackage{subcaption}
\usepackage{amsmath}
\usepackage{txfonts}
\usepackage{fix2col}
\usepackage{ulem}

\usepackage{color}
\usepackage{colortbl} % SdM: this is great, but will likely not be allowed in the final version of A&A. Check the proofs in the end!!
\definecolor{pinegreen}{RGB}{1, 121, 111}
\definecolor{salmon}{RGB}{255,160,122}
\usepackage[breaklinks=true,colorlinks,citecolor=pinegreen]{hyperref}         %SdM: added breaklinks (does not compile without for me) and backref (cause its neat)

\usepackage[hyphenbreaks]{breakurl}
\usepackage{natbib}

\definecolor{c1}{RGB}{91, 44, 111}
\definecolor{c2}{RGB}{13, 71, 161}
\definecolor{c3}{RGB}{14, 102, 85}

\definecolor{ylvacolour}{rgb}{.55,.4,.75}
\definecolor{YC}{RGB}{115,80,185}

\newcommand{\Msun}{\ensuremath{\,M_\odot}\xspace}

\newcommand{\kms}{\ensuremath{\,\rm{km}\,\rm{s}^{-1}}\xspace}
\newcommand{\Msunyr}{\ensuremath{\,M_\odot\,\rm{yr}^{-1}}\xspace}

\newcommand{\Iion}{\ensuremath{\,{I}_{\rm{ion}}}\xspace}
\newcommand{\Iionz}{\ensuremath{\,{I}_{\rm{ion, \ion{H}{I}}}}\xspace}
\newcommand{\Iiono}{\ensuremath{\,{I}_{\rm{ion, \ion{He}{I}}}}\xspace}
\newcommand{\Iiont}{\ensuremath{\,{I}_{\rm{ion, \ion{He}{II}}}}\xspace}
\newcommand{\ndotion}{\ensuremath{\,\dot{n}_{\rm{ion}}}\xspace}
\newcommand{\ndotionz}{\ensuremath{\,\dot{n}_{\rm{ion, \ion{H}{I}}}}\xspace}

\newcommand{\ndotiont}{\ensuremath{\,\dot{n}_{\rm{ion, \ion{He}{II}}}}\xspace}
\newcommand{\fesc}{\ensuremath{\,f_{\rm{esc}}}\xspace}
\newcommand{\fescm}{\ensuremath{\,f_{\rm{esc, massive}}}\xspace}
\newcommand{\fescs}{\ensuremath{\,f_{\rm{esc, stripped}}}\xspace}

\newcommand{\HI}{\ensuremath{\,\ion{H}{I}}\xspace}
\newcommand{\HII}{\ensuremath{\,\ion{H}{II}}\xspace}
\newcommand{\HeI}{\ensuremath{\,\ion{He}{I}}\xspace}
\newcommand{\HeII}{\ensuremath{\,\ion{He}{II}}\xspace}
\newcommand{\HeIII}{\ensuremath{\,\ion{He}{III}}\xspace}
\newcommand{\OIII}{\ensuremath{\,\ion{O}{III}}\xspace}
\newcommand{\CIII}{\ensuremath{\,\ion{C}{III}}\xspace}
\newcommand{\CIV}{\ensuremath{\,\ion{C}{IV}}\xspace}
\newcommand{\SiIV}{\ensuremath{\,\ion{Si}{IV}}\xspace}
\newcommand{\SiIII}{\ensuremath{\,\ion{Si}{III}}\xspace}

\newcommand{\code}[1]{{\textsc{#1}}}

\definecolor{midblue}{RGB}{44,133,255}

\newcommand{\struc}[1]{}
\newcommand{\conc}[1]{{}}

\newcommand{\edit}[1]{{\color{black} #1}}

\definecolor{darkcyan}{rgb}{0.0, 0.55, 0.55}
\definecolor{amethyst}{rgb}{0.6, 0.4, 0.8}
\usepackage{refs}   % Mine and Kibbi's package on how to refer to figures, equations, tables etc in A&A
\usepackage{booktabs}    % Nice tables

\defcitealias{2017A&A...608A..11G}{Paper~I}
\defcitealias{2018A&A...615A..78G}{Paper~II}
\defcitealias{2019A&A...629A.134G}{Paper~III}
\defcitealias{2014ARA&A..52..415M}{MD14}
\defcitealias{2012ApJ...746..125H}{HM12}

\usepackage{morefloats}

\bibpunct{(}{)}{;}{a}{}{,}

\begin{document} 

\title{The contribution from stars stripped in binaries to cosmic reionization of hydrogen and helium}
\titlerunning{Stripped stars during the reionization of hydrogen and helium}

   \author{Y.~G\"{o}tberg$^{1,3}$, S.~E.~de~Mink$^{2,3,4}$, M.~McQuinn$^{5}$, E.~Zapartas$^{6,3}$, J.~H.~Groh$^{7}$, \and C.~Norman$^{8,9}$}
   \authorrunning{G\"{o}tberg et al.}

   \institute{
            The Observatories of the Carnegie Institution for Science, 813 Santa Barbara St., Pasadena, CA 91101, USA \\
             \email{ygoetberg@carnegiescience.edu}
        \and
            Center for Astrophysics, Harvard \& Smithsonian, 60 Garden Street, Cambridge, MA 02138, USA
         \and
            Anton Pannekoek Institute for Astronomy, University of Amsterdam, 1090 GE Amsterdam, The Netherlands
         \and  
             GRAPPA, GRavitation and AstroParticle Physics Amsterdam, University of Amsterdam, 1090 GE Amsterdam, The Netherlands 
         \and 
            Department of Astronomy, University of Washington, Seattle, WA 98195, USA
        \and
            Geneva Observatory, University of Geneva, CH-1290 Sauverny, Switzerland
        \and
            School of Physics, Trinity College Dublin, The University of Dublin, Dublin 2, Ireland
        \and
            Department of Physics \& Astronomy, Johns Hopkins University, Baltimore, MD 21218, USA
        \and 
            Space Telescope Science Institute, 3700 San Martin Drive, Baltimore, MD 21218, USA
        }

%   \date{Received ......; accepted ......}

% ABSTRACT
\abstract
{
\edit{
Massive stars are often found in binary systems and it has been argued that binary products boost the ionizing radiation of stellar populations. Accurate predictions for binary products are needed to understand and quantify their contribution to Cosmic Reionization. 

We investigate the contribution of stars stripped in binaries since (1) they are, arguably, the best-understood products of binary evolution, (2) we recently produced the first non-LTE radiative transfer calculations for the atmospheres of these stripped stars that predict their ionizing spectra, and (3) they are very promising sources since they boost the ionizing emission of stellar populations at late times. This allows stellar feedback to clear the surroundings such that a higher fraction of their photons can escape and ionize the intergalactic medium.

Combining our detailed predictions for the ionizing spectra with a simple cosmic reionization model, we estimate that stripped stars contributed tens of percent of the photons that caused cosmic reionization of hydrogen, depending on the assumed escape fractions. More importantly, stripped stars harden the ionizing emission. We estimate that the spectral index for the ionizing part of the spectrum can increase to $-1$ compared to $\lesssim-2$ for single stars. At high redshift, stripped stars and massive single stars combined dominate the \HeII-ionizing emission, but we expect active galactic nuclei drive cosmic helium reionization. 

Further observational consequences we expect are (1) high ionization states for the intergalactic gas surrounding stellar systems, such as \CIV and \SiIV and (2) additional heating of the intergalactic medium of up to a few thousand Kelvin. Quantifying these warrants the inclusion of accurate models for stripped stars and other binary products in full cosmological simulations. 
}

}

   \keywords{binaries: close -- reionization -- galaxies: stellar content -- ultraviolet: stars}

   \maketitle

\section{Introduction}

% Which were the sources of cosmic reionization of hydrogen and helium? 
%   - patchy phase change of the Universe, used to probe cosmological evolution, crucial for understanding of XXX
%   - common consensus: stars did HI and quasars did HeII  (HeI together with HI for geometrical reasons)
%   - why is there a debate?: (1) cannot probe HeII-ionizing photons well, (2) measurements only available up to z~6 or so, while H-reionization occurred before that, (3) escape fraction needs to be high or cosmic SFH needs to be high at high z for H-reionization to be possible -- maybe not realistic
%   - why we even should care which were the sources?: constraints on the models, different spectral hardness, understanding of un-observable early Universe? 
Cosmic reionization refers to the process in the history of our Universe when the intergalactic medium (IGM) became ionized. How this process occurred is still uncertain and depends on two main questions: which were the sources of ionizing photons and what fraction of these photons escaped to become available for the ionization of the IGM.
%What the sources were that reionized the Universe is a long-standing question. %Its answer would reveal physical properties of the environment during the early Universe, which is crucial for understanding why the Universe looks like it does today \citep[e.g.,][]{2010Natur.468...49R}.
For the sources that emitted the ionizing radiation, which are the focus of this paper, the current consensus is that stellar populations produced most of the photons that ionized intergalactic hydrogen and neutral helium, while active galactic nuclei (AGN) produced the majority of the photons that fully ionized helium \citep{2001PhR...349..125B, 2009ApJ...703.1416F, 2016ApJ...825..144W, 2016ARA&A..54..313M}. However, the relative contributions from AGN and stellar populations have not been fully constrained. % and more detailed studies of the stellar types that provide the ionizing emission from stellar populations is needed.

% cosmological simulations with stars -- great advances 
%   - manage to model reionization
%   - need high escape fractions
%     Introduce: escape fraction (fesc), intergalactic medium (IGM)
Numerical and analytical models of the early Universe \citep[e.g.,][]{1997ApJ...486..581G, 2007MNRAS.377.1043M, 2010Natur.468...49R, 2012ApJ...746..125H, 2018MNRAS.480.2628F, 2019MNRAS.483.1029K} reproduce the observed completion of hydrogen reionization around a redshift, $z$, of $\sim 6$ \citep{2006AJ....132..117F}. % and of helium reionization around \citep{z \sim 3} \citep[e.g.,][]{2008ApJ...681....1F}. 
However, to achieve this, the models require that a large fraction of the produced ionizing photons escape the host galaxies and reach the IGM. Only when assuming a high escape fraction (\fesc) of 20\% or more at high redshifts, the simulations reach complete reionization by $z\sim 6$ as observations indicate \citep{2007MNRAS.382..325B, 2009ApJ...706.1136O, 2012ApJ...746..125H, 2013ApJ...768...71R}. With a low escape fraction of a few percent, similar to what is observed locally and at intermediate redshifts \citep[e.g.,][]{2016Natur.529..178I,2018ApJ...869..123S}, the reionization is significantly delayed, which does not match several observational constraints \citep{2006ARA&A..44..415F}. 
While it is possible that the escape fraction indeed was higher in the early Universe \citep[][]{2006MNRAS.371L...1I}, other effects could impact the amount of ionizing photons that reach the IGM at high redshifts. Examples are higher cosmic star-formation rate than usually assumed because of faint and therefore un-detected galaxies in the distant Universe \citep[see][]{2015ApJ...814...69A, 2017ApJ...835..113L} and higher ionizing emission from quasars at high redshift than usually considered \citep{2015ApJ...813L...8M}. Here we explore additional contribution from stellar sources that have not been investigated in detail; stars stripped of their hydrogen-rich envelopes in binary interaction with a close companion.

% % % %   On the missing ionizing stars   % % % % - can I make a list? maybe that is to teaching-like... 
% First paragraph - launch the concept!
%   - main message: ionizing sources are neglected, since usually just main-sequence stars and WR stars are considered. Let's describe these sources!
%   - recent advance on accounting for rotation and higher mass stars - however rotation, just one rotation speed, not realistic, and VMSs we cannot really model because we don't know much about wind mass-loss
%   - Stars are in binaries that are so close that they inevitable will interact during their lifetime (Sana, Moe)
Recent advances in stellar astrophysics indicate that several types of stars that emit ionizing radiation are still often neglected in models of the ionizing output from galaxies. Efforts have been made to include stellar rotation \citep{2005A&A...429..581M, 2012ApJ...751...67L, 2013A&A...558A.103G, 2014ApJS..212...14L}, and very massive stars with masses above 100\Msun \citep[][]{2016MNRAS.458..624C} in models for the spectra of stellar populations. Observational surveys show evidence that massive and intermediate mass stars have binary companions so close that interaction between the two stars is inevitable as the stars swell during their evolution \citep[e.g.,][]{2012Sci...337..444S, 2017ApJS..230...15M}.
Such binary interaction can lead to severe exchange or loss of stellar material and possibly even coalescence of the stars. %Many of the resulting binary products are thought to increase the number of emitted ionizing photons \citep[e.g.,][]{1999NewA....4..173V}. 

% Add credit to BPASS here
\edit{Pioneering work has been done by the team behind the binary population and spectral synthesis (\code{BPASS}) code \citep{2009MNRAS.400.1019E, 2012MNRAS.419..479E, 2017PASA...34...58E}, who have provided observationally testable predictions for stellar populations that account for binary interaction. These include for example emission rates of ionizing photons \citep[][see also \citealt{1999NewA....4..173V}]{2016MNRAS.456..485S}, the production efficiency of H-ionizing photons \citep[$\xi_{\text{ion}}$,][]{2016MNRAS.458L...6W}, and nebular emission line ratios for nearby and distant stellar populations \citep{2014MNRAS.444.3466S, 2018MNRAS.477..904X}. Using these results, sophisticated models including the effect of interacting binaries have been made for cosmic reionization and the evolution of galaxies  \citep{2016MNRAS.459.3614M, 2018MNRAS.479..994R}. %These works showed how interacting binaries give rise to reionization at about the predicted time with realistic assumptions for the escape fraction.
These works showed that interacting binaries give rise to reionization earlier than what is predicted by stellar populations containing only single stars. 
They have also shown that interacting binaries may result in additional heating of the IGM, and that the delay in the production of ionizing photons from interacting binaries relative to isolated massive stars may allow for higher escape fractions.} 

% Why that is not sufficient
\edit{However, disentangling the different contributions from individual sources is necessary to both understand which stars that are responsible for the ionizing emission and, maybe more importantly, how uncertain the predicted emission rates of ionizing photons are. Here, we have, therefore, chosen to focus on modeling the contribution from one type of binary product: stars stripped of their hydrogen-rich envelopes. 
% Focus on stripped stars
% - what are they?
% - why their ionizing radiation is interesting
Such} envelope-stripping is predicted to be the most common type of interaction in binaries as it is the fate of a third of all massive stars \citep{2012Sci...337..444S}. During either stable or unstable mass-transfer, the hydrogen-rich envelope of one star is lost or partially transferred to the second star, leaving the hot, helium-rich core exposed \citep{1967ZA.....65..251K, 1967AcA....17..355P, 1992ApJ...391..246P, 2013A&ARv..21...59I}. These stripped stars are so hot ($\sim 50-100$~kK) that the majority of their emitted photons are able to ionize hydrogen \citep[][hereafter \citetalias{2017A&A...608A..11G}]{2017A&A...608A..11G}. The envelope-stripping is \edit{usually completed after or once the donor star reaches central hydrogen exhaustion}, meaning that stripped stars are formed with a time delay that roughly corresponds to the main-sequence duration of the donor star. Because stars over a wide range of masses can become stripped, this means that the ionizing radiation from stripped stars is emitted over an extended period of time after a starburst \citep[several $100$~Myr,][]{2019A&A...629A.134G}. It is likely that the ionizing radiation from stripped stars therefore escapes the birth environment and reaches the IGM, as strong stellar winds and the first supernovae are expected to remove most of the surrounding gas within several Myr after star formation started \citep[e.g.,][]{2018arXiv181201615K}. \edit{In the case some gas is still present when stripped stars are created, it is likely that their hard ionizing emission produces characteristic nebular emission line morphology \citep[][see also \citealt{2014MNRAS.444.3466S}]{2018MNRAS.477..904X}.}

% Proof that stripped stars exist
Envelope-stripping is probably also the type of binary interaction that is most well-understood, both because models suggest that it is a common and often inevitable evolutionary phase and also because several stripped stars have been observed \citep{1998ApJ...493..440G, 2008A&A...485..245G, 2008ApJ...686.1280P, 2013ApJ...765....2P, 2017ApJ...843...60W, 2018ApJ...853..156W, 2018ApJ...865...76C}. The number of observed stripped stars is low, but the reason is likely that most stripped stars are hidden by their bright companion stars in optical wavelengths and therefore difficult to discover \citepalias[][see also \citealt{2018ApJ...853..156W}]{2017A&A...608A..11G}. The frequently observed hydrogen-poor supernovae are also evidence for the existence of a large population of stripped stars \citep{2011MNRAS.412.1522S, 2016MNRAS.457..328L, 2016ApJ...827...90L, 2017PASP..129e4201S, 2017ApJ...837..121G}. Their low ejecta masses \citep[e.g.,][]{2011ApJ...741...97D, 2016ApJ...832..108M} and the apparent difficulty to directly observe their progenitors \citep[e.g.,][]{2013MNRAS.436..774E} suggest that most of these explosions are not the deaths of massive Wolf-Rayet stars, but originate from lower mass, hydrogen-poor stars\edit{, which is also in agreement with the old ages inferred for hydrogen-poor supernova sites using nebular emission features \citep{2019MNRAS.482..384X}}.
Additionally, in the classical framework of isolated binary evolution, stripped stars are thought to play important roles as the progenitors of gravitational wave events \citep{2003MNRAS.344..629D, 2016Natur.534..512B, 2017ApJ...846..170T, 2018MNRAS.481.1908K}. With the rapidly increasing number of detected gravitational wave events, understanding the impact of stripped stars both nearby and over cosmic time is becoming urgent.  
%Another observational constraint is the increasing number of observed gravitational wave events, which are thought to have stripped stars as progenitors in the isolated binary evolution channels \citep[e.g.,][]{2003MNRAS.344..629D, 2016Natur.534..512B, 2017ApJ...846..170T, 2018MNRAS.481.1908K}. %Stripped stars that are sufficiently massive are thought to explode in stripped-envelope supernovae, which are frequently observed \citep{2016MNRAS.457..328L, 2016ApJ...827...90L, 2017PASP..129e4201S, 2017ApJ...837..121G} and thus an additional evidence of the existence of stripped stars \citep{2011MNRAS.412.1522S}. Stripped stars also constitute a necessary step towards the merger of two neutron stars \citep{2003MNRAS.344..629D, 2017ApJ...846..170T, 2018MNRAS.481.4009V}, which recently was detected \citep{2017PhRvL.119p1101A}. 

% Describe briefly the other stellar sources of ionizing emission
%With this as motivation, we here focus on the contribution from stripped stars to the budget of photons that reionized the Universe. We note that there are several other binary products that likely contributed, but which are not the topic of this work. 
\edit{In addition to stripped stars, there are several other binary products that likely contributed to the photon budget during the reionization, but which are not the topic of this work. }
% Mass gainers and mergers (rejuvenation)
For example, stars that gain mass through mass transfer with a companion star and binary stars that merge. These stars are thought to appear as massive blue stragglers, i.e., more massive main sequence stars compared to their progenitors, leading to an increase of the ionizing emission from massive OB type main sequence stars \citep{1995A&A...297..483B, 1999NewA....4..173V,2003A&A...400..429B, 2009MNRAS.395.1822C,2014ApJ...780..117S, 2014ApJ...782....7D}.
% CHE (both via MT and tidal forces, primarily low Z?)
Another effect of binary interaction is rapid rotation induced by either mass accretion, coalescence, or tidal forces, which can cause the interior of the rotating stars to mix, providing the burning regions with fresh fuel \citep{1981A&A...102...17P, 2009A&A...497..243D, 2013ApJ...764..166D}. If wind mass-loss and the associated loss of angular momentum is low, as is the case at low metallicity, and depending on the properties of internal magnetic fields and the efficiency of rotational mixing, the star continues to rotate rapidly and may evolve chemically homogeneously \citep{2005A&A...443..643Y, 2007A&A...465L..29C, 2008MNRAS.384.1109E,2009A&A...497..243D, 2019A&A...627A..24G}. The result is that the stars remain small and hot throughout their lives, keeping their ionizing emission rates high \citep{2015A&A...581A..15S, 2019A&A...623A...8K}.
% Accreting compact objects (XRBs and accreting WDs)
Lastly, accreting white dwarfs, neutron stars, or black holes are expected to radiate X-rays and dominate the output of 1~keV photons from stellar populations but are likely unimportant sources of the extreme ultraviolet photons responsible for the reionization of hydrogen, but potentially interesting for helium reionization \edit{ \citep[e.g.,][see also \citealt{2013MNRAS.432.1640W, 2014MNRAS.439.2351W, 2019arXiv190910574S}]{2013ApJ...764...41F, 2015MNRAS.453.3024C, 2017ApJ...840...39M, 2019A&A...622L..10S}.} 
%White dwarfs accreting material have softer emission than accreting neutron stars or black holes and could be significant contributors of \HeII-ionizing radiation \citep{2015MNRAS.453.3024C, 2019A&A...622L..10S}, although 
However, calculations suggest that accreting compact objects are likely subdominant to quasars \citep{2018ApJ...869..159U}.

\edit{Well-established spectral synthesis codes that commonly are used to represent the radiative emission from stars during reionization are, for example, \code{Starburst99} \citep{1999ApJS..123....3L, 2014ApJS..212...14L} and \code{GALAXEV} \citep{2003MNRAS.344.1000B}.} %Efforts have been made to account for the effect of interacting binaries \citep{1998A&A...334...21V,  2004A&A...415..117Z, 2007ApJ...662L.107V, 2009MNRAS.395.1822C}, and with the progress made by \code{BPASS} an increasing number of cosmological simulations account for a variety of binary products \citep[e.g.,][]{2016MNRAS.458L...6W, 2018MNRAS.479..994R}.}
Efforts have been made to account for the effect of interacting binaries \citep{1998A&A...334...21V,  2004A&A...415..117Z, 2007ApJ...662L.107V, 2009MNRAS.395.1822C}, \edit{with particular progress in the field of the distant Universe made by \code{BPASS} as previously mentioned. }
%
% The radiative emission from stars is usually represented with spectral synthesis codes, which combine the evolutionary and spectral models of stars \citep[for a review, see][]{2013ARA&A..51..393C}. Established spectral synthesis codes have been tested to model the contribution from single stars well \citep[e.g., \code{Starburst99} and \code{GALAXEV},][]{1999ApJS..123....3L, 2014ApJS..212...14L, 2003MNRAS.344.1000B}. Efforts have been made to account for the effect of interacting binaries \citep{1998A&A...334...21V,  2004A&A...415..117Z, 2007ApJ...662L.107V, 2009MNRAS.395.1822C}, and recent advances have been made with the code \code{BPASS} \citep{2009MNRAS.400.1019E, 2012MNRAS.419..479E, 2017PASA...34...58E}. %, which provides testable predictions for a variety of observational phenomena \citep{2013MNRAS.436..774E, 2014MNRAS.444.3466S, 2016MNRAS.456..485S}.
% % Small discussion about BPASS
% In \code{BPASS}, a variety of binary products are accounted for, which together result in more ionizing photons being emitted when binaries are included and the emission rates of ionizing photons are also higher at late times \citep{2016MNRAS.456..485S, 2016MNRAS.459.3614M}. Disentangling the different contributions from individual sources is necessary to both understand which stars are responsible for the ionizing emission and, maybe more importantly, how uncertain the predicted emission rates of ionizing photons are.  

Modeling the radiative contribution from stripped stars has been a challenge since no spectral models made for stripped stars have been available, and their spectra have been represented with spectral models made for other types of stars \citep[e.g.,][]{2017PASA...34...58E}. However, in \citet[][hereafter \citetalias{2018A&A...615A..78G}]{2018A&A...615A..78G} we presented grids of evolutionary and spectral models made for stars stripped in binaries, which we synthesized to estimate the contribution from stripped stars to stellar populations in \citet[][hereafter \citetalias{2019A&A...629A.134G}]{2019A&A...629A.134G}. These models are publicly available on CDS\footnote{\url{http://cdsweb.u-strasbg.fr/cgi-bin/qcat?J/A+A/}} and via the \code{Starburst99} online interface\footnote{\url{http://www.stsci.edu/science/starburst99/docs/default.htm}}.

% Aim of this study: Estimate the importance of stripped stars during reionization of hydrogen and helium/over cosmic history
%   - simple/analytical model is the best to start with since it is then easier to see actually the effects that stripped stars give and not effects related to something more complex
%   - we do this unique work, by for the first time using detailed spectral models for the stripped stars using the recent update of Starburst99
%   - Fourth paper in a series on the role of ionizing radiation from stripped stars, here focusing on large-scale implications
In this paper, we estimate the importance of stars stripped in binaries over cosmic history and discuss their roles during the reionization of hydrogen and helium. This is the fourth paper in a series in which we describe and discuss the impact of the ionizing radiation emitted by stripped stars, however, it can be read independently. We use our \edit{evolutionary and spectral models made with the stellar evolution code \code{MESA} \citep{2011ApJS..192....3P,2013ApJS..208....4P,  2015ApJS..220...15P, 2018ApJS..234...34P, 2019ApJS..243...10P} and the radiative transfer code \code{CMFGEN} \citep{1990A&A...231..116H, 1998ApJ...496..407H}} presented in \citetalias{2018A&A...615A..78G} and \edit{our population synthesis presented in} \citetalias{2019A&A...629A.134G} together with a simple model for cosmic reionization to understand the relative contributions from stripped stars during the evolution of the Universe. \edit{We emphasize that our novel approach of developing custom-made atmosphere models for stripped stars using CMFGEN allows us to study them in more depth and thus to better understand their role in stellar populations over cosmic time than previous studies.}

% Structure of the paper:
%   - Sect. 2: Description of the analytical model
%   - Sect. 3: How we describe the ionizing radiation from stars and quasars  (incl. fesc?)
%   - Sect. 4: Resulting reionization models and relative contributions from sources
%   - Sect. 5: Observable consequences of stripped stars
%   - Sect. 6: Summary & Conclusions
We structure the article as follows. 
In \secref{sec:model}, we describe how we model the ionizing radiation from stellar populations, from AGN, and over cosmic time. 
In \secref{sec:SED_results}, we present the ionizing emission and the influence of stripped stars on the hardness of the ionizing radiation that emerges into the IGM. 
In \secref{sec:reionization}, we describe the role of stripped stars during reionization by quantifying the fraction of ionizing photons they contribute with and what their effect is on when reionization is completed. 
Our models suggest that stripped stars impact the conditions in the IGM, which in some cases may influence observable quantities. In \secref{sec:discussion}, we therefore discuss the possible effect of stripped stars on intergalactic absorption features from metals and heating of the IGM. 
In \secref{sec:summary_conclusions}, we summarize our findings and conclusions.

%________________________________________________________________

\section{Modeling cosmic reionization}\label{sec:model}

\begin{figure}
\centering
\includegraphics[width=\hsize,trim=10cm 55mm 11cm 3cm,clip]{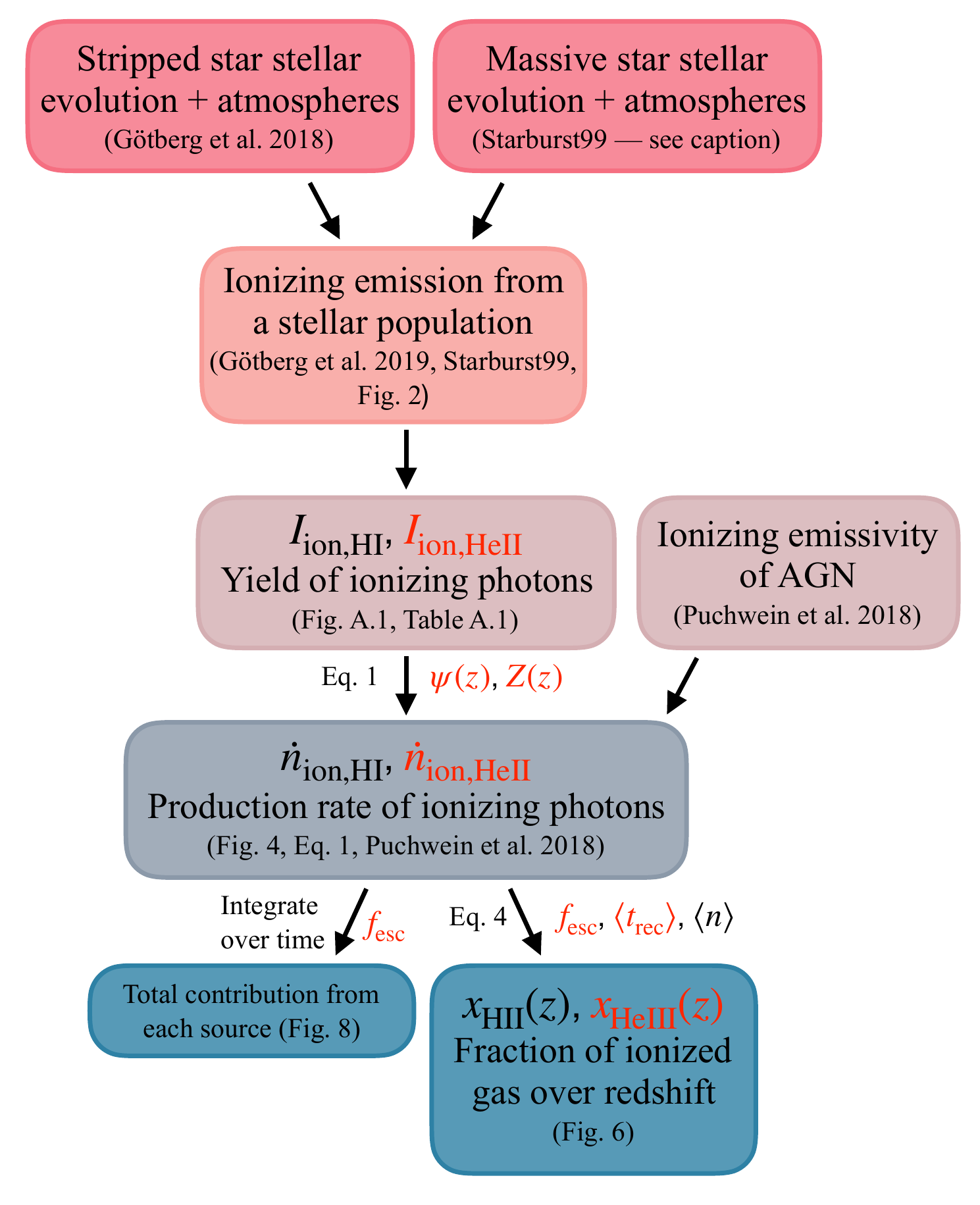}
\caption{Flowchart visualizing the method we use to model the reionization of hydrogen and helium accounting for ionizing radiation from both stellar populations and AGN. The quantities that we consider uncertain are written with red color. See \secref{sec:model} for details and also, for example, \citet{1999ApJ...514..648M} and \citet{2012ApJ...746..125H} where similar methods have been employed. (The massive stars in \code{Starburst99} are represented by evolutionary models from the Geneva grids \citep{2012A&A...537A.146E,2012ApJ...751...67L,2013A&A...558A.103G}, atmosphere models from \code{CMFGEN} for Wolf-Rayet stars \citep{1998ApJ...496..407H, 2002MNRAS.337.1309S}, and atmosphere models from \code{WM-Basic} for OB-stars \citep{2001A&A...375..161P, 2010ApJS..189..309L}.)}
\label{fig:flowchart}
\end{figure}

% What is happening in this section?
In this section, we first describe the models that we adopt for the sources of ionizing radiation in \secref{sec:ionizing_sources}. Then, we describe the semi-analytical approach we take to simulate the reionization of hydrogen and helium in \secref{sec:analytical_model}. We visualize the method in \figref{fig:flowchart}.

% This section should describe in detail the models that we use for the ionizing sources
% boing section, lots of details, no figures.
\subsection{Sources of ionizing radiation}\label{sec:ionizing_sources}

% Consider three sources of ionizing radiation over cosmic time: stripped stars, massive stars, and AGN
We consider stars stripped in binaries, massive main-sequence stars, Wolf-Rayet stars, and AGN as sources of ionizing radiation during the reionization of hydrogen and helium.
% Below, we give details for how we model the ionizing emissivity of the difference sources
Below, we describe how we model the ionizing emissivity over cosmic time for each of these sources.

% Stellar sources of ionizing radiation
\subsubsection{Stellar populations including stars stripped in binaries}\label{sec:stellar_populations}

% % % % %    Stripped stars    % % % % % 

\begin{figure}
\centering
\includegraphics[width=.9\hsize]{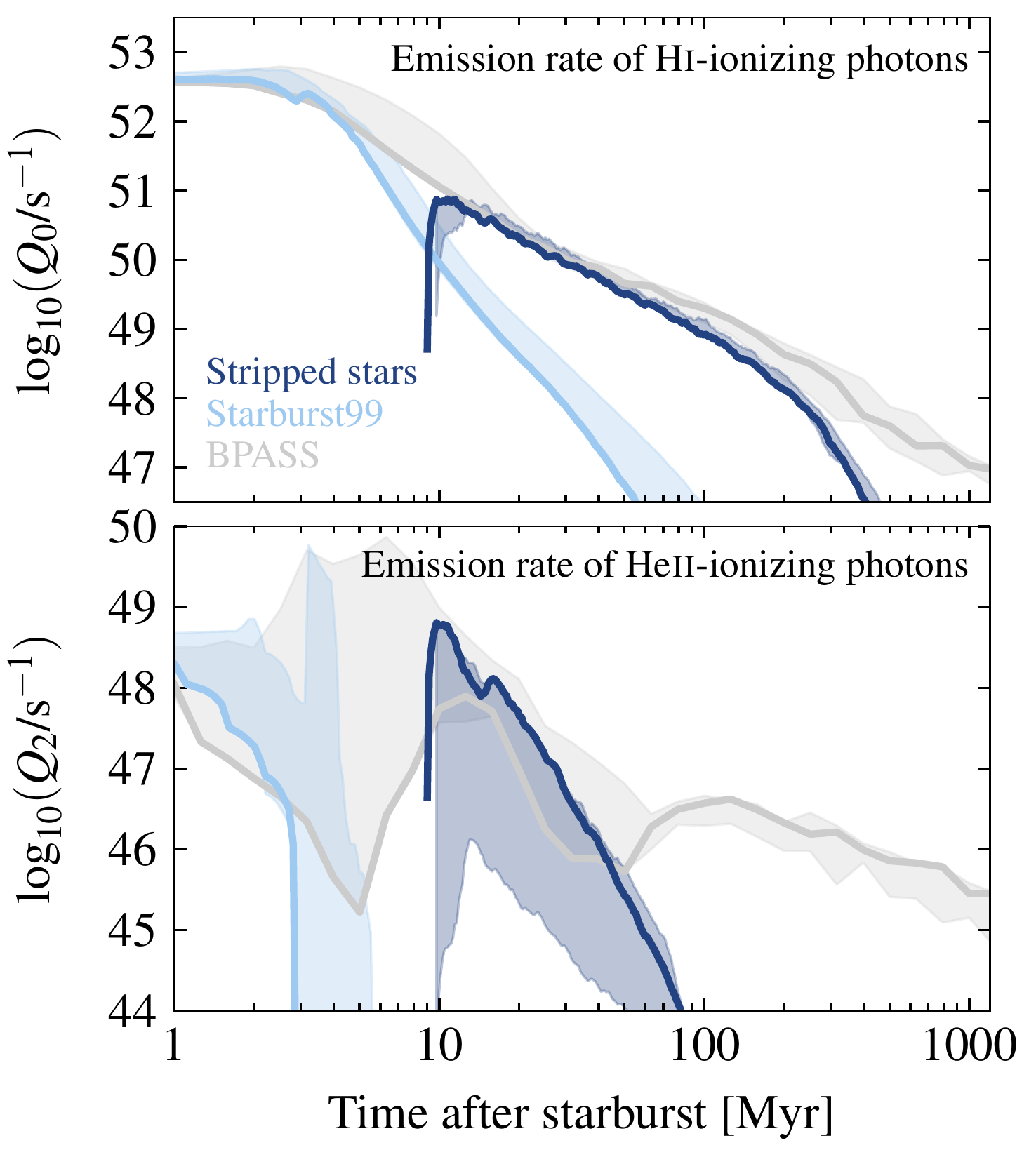}
\caption{The emission rates of ionizing photons from a starburst of $10^6$\Msun as a function of time. We show the emission rates from stripped stars in dark blue, from massive stars in light blue (using \code{Starburst99}), and we also show the predicted emission rates from the code \code{BPASS} in gray. The shaded regions span the emission rates for metallicities between $Z = 0.0002$ and $Z = 0.014$ and the solid lines show the emission rates for solar metallicity  ($Z = 0.014$). The top panel shows the emission rate of \HI-ionizing photons, while the bottom panel shows the emission rate of \HeII-ionizing photons. It is visible that stripped stars are responsible for late emission of ionizing photons in a starburst. This figure is adapted from \citetalias{2019A&A...629A.134G}.  %(This figure is adapted from \citet{2019A&A...629A.134G}.)
}
\label{fig:Qt}
\end{figure}

% Physics: Consider the contribution from stars stripped in binaries (case A, B, B/CEE)
We consider stripped stars that are created via stable mass-transfer and common envelope evolution. 
%We consider stars that are stripped in binaries via the three main formation channels. First, stable mass-transfer initiated during the main-sequence and Hertzsprung gap evolution of the donor star \citep[Case~A and B type mass-transfer, respectively,][see also \citealt{2002ApJ...573..283P}]{1967ZA.....65..251K}. Then, also, common envelope evolution that is initiated during the Hertzsprung gap evolution of the donor star and that is followed by a successful ejection of the envelope as the stars spiral closer together. 
% Characteristics of stripped stars: hot, long-lived, but can form with a delay (Paper I and II)
In these formation channels, the compact helium core of the donor star is exposed. This stripped star is hot ($\gtrsim 30\,000$~K) and long-lived if interaction occurred prior to the completion of central helium burning, which is an evolutionary phase that constitutes about 10~\% of the total stellar lifetime (see \citetalias{2017A&A...608A..11G} and \citetalias{2018A&A...615A..78G}). Stripped stars are formed over an extended time period after a burst of star-formation, mainly because they originate from lower mass stars than massive WR and O-type stars. This allows the ionizing radiation to be emitted up to several hundred Myr after a starburst \citepalias{2019A&A...629A.134G}. This is seen in \figref{fig:Qt}, where we show the emission rates of \HI- and \HeII-ionizing photons from a starburst. The massive stars modeled by \code{Starburst99} are predicted to emit ionizing photons early, primarily within the first $\sim 10$~Myr. Stripped stars are formed after $\sim 10$~Myr and emit ionizing photons for several hundred Myr. This is consistent with some of the predictions from the binary population synthesis code \code{BPASS}. For more details, see \citetalias{2019A&A...629A.134G} from which \figref{fig:Qt} is adapted.
 
% Use models for case B, but that is pretty good for all of these formation channels (models from Paper II)
We model the ionizing output from stripped stars using the detailed evolutionary and spectral models presented in \citetalias{2018A&A...615A..78G}. These models are custom-made for stars stripped in binaries. They were computed for initial donor star masses between 2 and 18\Msun using the binary stellar evolutionary code \code{MESA} \edit{\citep[][]{2011ApJS..192....3P, 2013ApJS..208....4P, 2015ApJS..220...15P, 2018ApJS..234...34P, 2019ApJS..243...10P}} and the non-LTE radiative transfer code \code{CMFGEN} \citep{1990A&A...231..116H, 1998ApJ...496..407H}.
%The models are made for envelope-stripping via Case~B type mass-transfer initiated early during the Hertzsprung gap evolution of the donor star. We expect that the large-scale difference between these models and the products from Case~A type mass-transfer or through common envelope evolution are negligible. The largest effects are for Case~A type mass-transfer, which results in lower-mass stripped stars compared to Case~B type mass-transfer \citep{1994A&A...290..119P}. 
%Additionally, the initial period has an effect on the amount of envelope that is left after mass-transfer. The amount affects the radius evolution of the star slightly and sometimes significantly, for example, at low metallicity \citep{2011A&A...528A.131C, 2017ApJ...840...10Y}.
We consider four metallicities $Z = 0.014$, 0.006, 0.002, and 0.0002, which correspond roughly to the metallicity of the Sun, the Large and Small Magellanic Clouds, and a very low metallicity environment that likely existed at high redshifts.

% Synthesize the detailed evolutionary and spectral models into a population assuming IMF, fbin, P distr, q distr, and using single stellar evolutionary models to get interaction, then assume qcrit for outcome of interaction (together with the alpha prescription). This is detailed in Paper III.
We compute the total radiative output from stripped stars by simulating a stellar population based on our detailed models. Details can be found in \citetalias{2019A&A...629A.134G}, but we summarize the main assumptions here. We draw initial stellar masses, $M_{\text{init}}$, assuming a \citet{2001MNRAS.322..231K} initial mass function (IMF) with lower and upper mass limits at $0.1$ and $100$~\Msun. We determine which stars are the most massive stars in binary systems by assuming the mass-dependent binary fraction of \citet{2017ApJS..230...15M}. We draw companion stars to the primaries by assuming a flat mass ratio ($q \equiv M_{\text{init, 2}}/M_{\text{init, 1}}$) distribution between 0.1 and 1, in agreement with observations from \citet{2012ApJ...751....4K}, \citet{2012Sci...337..444S}, and as summarized in \citet{2017ApJS..230...15M}. We assume that the initial orbital period for the binary systems follows the distribution of \citet{2012Sci...337..444S} for binaries in which the most massive star in the system is more massive than 15\Msun and the distribution of \citet{1924PTarO..25f...1O} in case it has a lower mass than 15\Msun. We assume the initial period range to span from binaries that touch at birth up to $10^{3.7}$~days \citep{2017ApJS..230...15M}. We determine when interaction is initiated by comparing the sizes of the Roche-lobes of the stars \citep{1983ApJ...268..368E}, with the sizes of the stars, which we measure from evolutionary models of single stars that have the same assumptions for the interior as our binary models \citepalias{2018A&A...615A..78G}. Whether the system initiates stable mass-transfer or common envelope evolution is determined by comparing the mass ratio to a critical value, $q_{\text{crit}}$, above which stable mass-transfer follows and below which a common envelope develops ($q_{\text{crit}} = 0.65$ and 0.4 for main-sequence and Hertzsprung gap donors, following \citealt{2007A&A...467.1181D} and \citealt{2013ApJ...764..166D}, respectively). In case of stable mass-transfer, the result is a binary system containing a stripped star. For common envelope, we determine whether the envelope is successfully ejected by taking the classical $\alpha$-prescription \citep{1984ApJ...277..355W} \edit{and setting the efficiency parameter, $\alpha_{\text{CE}}$, and the structure parameter, $\lambda_{\text{CE}}$, to be $\alpha_{\text{CE}}\lambda_{\text{CE}} = 0.5$, which is comparable with current literature \citep[e.g.,][]{2000A&A...360.1043D,2002MNRAS.329..897H, 2004PhDT........45I, 2019arXiv190712573F}. This choice results in only few stripped stars created via common envelope ejections in our models \citepalias{2019A&A...629A.134G}, but we note that the common envelope parameters are poorly constrained, which could affect the predictions for the number of stripped stars in stellar populations.} We assume that all stripped stars with progenitor stars of the same mass have the same properties, independent on the formation channel. 

% Model uncertainties for stripped stars
%   - ionizing emission from stripped stars is a remarkably robust prediction from models
%   - binary fraction is well-constrained, mass ratio doesn't matter much, period distribution doesn't matter much
%   - outcome of binary interaction appears robust from models -- all evolutionary models with same initial conditions produce stripped stars
%   - wind mass-loss should be low, and therefore is the H and HeI-ionizing emission quite robust -- HeII ionizing emission sensitive as it is in the Wien part
We consider our predicted emission rates of \HI- and \HeI-ionizing photons from stripped stars to be relatively robust, since the properties of stripped stars are predicted to be insensitive to the orbital period or the assumed mass ratio in most cases \citep[see however][\edit{who showed that the properties of stripped stars are affected at very low metallicities or at very long initial periods}]{2017ApJ...840...10Y, 2018arXiv180807580S}. The number of stripped stars present in stellar populations also impacts the emission rates of ionizing photons. We consider the predicted number of stripped stars to be relatively robust since initial conditions for massive binaries, such as the binary fraction, orbital period distribution and mass ratio distribution, have been constrained observationally \citep[][see also \citealt{2017ApJS..230...15M}]{2012Sci...337..444S}. Using binary population synthesis models together with observed data \citet{2012Sci...337..444S} found that 1/3 of massive stars become stripped. 
%We consider our predicted emission rates of \HI- and \HeI-ionizing photons from stripped stars to be relatively robust, since great efforts have been made to constrain the properties of young massive binaries. When observing massive stars in Galactic clusters, \citet{2012Sci...337..444S} found that $0.69\pm 0.09$ of stars are binaries, which also preferentially have short periods. Combining binary population synthesis with observations, \citet{2012Sci...337..444S} found that about 33\% of massive stars become stripped. The properties of stripped stars are also predicted to be insensitive to the initial orbital period or the assumed mass ratio in most cases \citep[see however][]{2017ApJ...840...10Y, 2018arXiv180807580S}. 
Stellar wind mass-loss from stripped stars is not well-known, which is important to take into account since dense stellar winds can block ionizing emission. In \citetalias{2017A&A...608A..11G}, we found, however, that \HI- and \HeI-ionizing emission from stripped stars only varies by a factor of two for a range of wind mass loss rates. In contrast, the \HeII-ionizing emission is sensitive to variations in the wind mass-loss rate, and we therefore consider our predicted emission rates of \HeII-ionizing photons to be less certain \citepalias[see][for a discussion]{2017A&A...608A..11G}. However, as we show in this study, the contribution from stripped stars to the ionization of \HeII\ is anyway much smaller than that from quasars.

% Additional comments on uncertainties
Our models concern only stripped stars with progenitors of initial masses $\leq 20 \Msun$, which excludes the contribution from higher mass stripped stars that could provide an extra boost of ionizing emission. Because it is uncertain how common such massive stripped stars are and how hot they are, predictions for their contribution are also uncertain. 
In addition, our models for stripped stars are made for stable mass-transfer, which likely leads a small amount of the hydrogen-envelope to remain in contrast to what is expected from the violent common envelope evolution (cf.\ \citealt{2011ApJ...730...76I}; \citealt{2017ApJ...840...10Y}; \citetalias{2017A&A...608A..11G}). Because the stripped stars created through common envelope evolution most likely have less of they hydrogen envelope left after interaction, they are expected to be hotter, which also leads them to emit harder ionizing radiation.
The difference in ionizing emission rates between the stripped stars created via the two evolutionary channels is small at high metallicity, but it could be significant for low-metallicity environments \citepalias[cf.][]{2019A&A...629A.134G}. We expect that including more accurate models for stripped stars created through common envelope evolution would only weakly affect the emission rate of \HI-ionizing photons, but more significantly the emission rate of \HeII-ionizing photons. The reason is that the bulk of the emission from our stripped star models is already \HI-ionizing, but the \HeII-ionizing emission is sensitive to temperature variations. 
To accurately account for the ionizing emission from high-mass stripped stars and the higher temperatures from stripped stars created through common envelope evolution, more detailed evolutionary models are required. Such studies is important, but out of the scope of this paper.

% % % % %    Massive stars    % % % % % 

% Can be combined with models for massive stars to represent the total output from stellar populations
To represent the ionizing output from a full stellar population including massive main sequence stars and Wolf-Rayet stars, we combine our predictions for stripped stars with estimates for single stars using the spectral synthesis code \code{Starburst99} \citep{1999ApJS..123....3L, 2010ApJS..189..309L, 2014ApJS..212...14L} as detailed in \citetalias{2019A&A...629A.134G}. %We use the same initial mass function as for the population synthesis described above, and assume the metallicities $Z = 0.014$, 0.008, 0.002, and 0.001 to combine with the models for stripped stars.
Binaries that have not yet interacted are mostly main sequence stars, and can therefore also be considered to be well-modeled by \code{Starburst99}. Our approach does not account for the emission from other binary products, such as merger stars or stars that gained mass through accretion.

% Compared to BPASS - what are the differences? 
%   - We have first accurate model for stripped stars -- can single out their contributions
%   - we neglect other types of binary products, which may play a role
%   - BPASS takes a more global approach and includes more different types of binary products
%   - however, they do not have detailed spectral models for stripped stars and they account for uncertain binary interaction channels (rotational mixing) 
%We use spectral models especially made for stripped stars in our models, which we expect will provide more accurate predictions than when down-scaled WR star models or blackbody spectra are used to represent the radiation from stripped stars as in earlier spectral synthesis models. These detailed models allow us to study the contribution from stripped stars in detail. However, we do not account for other types of binary products, some of which could contribute with additional ionizing radiation \citep[cf.][for further discussion]{2016MNRAS.456..485S, 2017PASA...34...58E, 2018MNRAS.477..904X}. 
%The code \code{BPASS} includes models for several different binary products, including uncertain predictions from rotational mixing and chemically homogeneous evolution. The representation of stripped stars is simplified, using spectral models made for WR stars, but agrees with our models \citepalias[see][]{2019A&A...629A.134G}. 

% % % % %   How to get the yield of ionizing photons I_ion   % % % % % 

\paragraph{Yields of ionizing photons, \Iion} ~\\
%
% How to quantify the emission of ionizing photons from stellar populations? integrate Q over time -> Iion
%   over cosmic time, this is an ok assumption because the Universe is much older than time it takes to make photons, apart from at really high redshifts (z > X)
\noindent A simple way to compare the ionizing radiation from different types of stars is to integrate their emission rates of ionizing photons (shown in \figref{fig:Qt}) over time and normalize by the total mass of formed stars. This quantity is referred to as the  yield of ionizing photons, \Iion, and describes the number of produced ionizing photons per solar mass of formed stars, or, equivalently, the emission rate of ionizing photons per star-formation rate \citep[see][]{2014ARA&A..52..415M}. %Using the yield of ionizing photons as a production rate is justified over cosmic history. However, since stars emit ionizing photons for an extended period in time, it is not valid to use \Iion as a production rate at very high redshift ($z \gtrsim 30$) when the emission delay is large compared to the age of the Universe \citepalias[see][]{2019A&A...629A.134G}. 

% What are the results? Can we refer back to the SED_sources?
For the yields of \HI-ionizing photons, we find that stripped stars produce $\Iionz \approx 10^{51.9}$ s$^{-1}$ (\Msun year$^{-1}$)$^{-1}$, while massive stars produce $\Iionz \approx 10^{53}$ s$^{-1}$ (\Msun year$^{-1}$)$^{-1}$ in the case of solar metallicity. This means that stripped stars are responsible for about 5\% of the total production of \HI-ionizing radiation from stellar populations. Despite providing a small fraction of the total budget, we argue that the delay in the photon production from stripped stars likely enhances their importance for reionization (see \secref{sec:reionization}). For the yield of \HeII-ionizing photons at solar metallicity, our models predict that $\Iiont \approx 10^{49.1}$ s$^{-1}$ (\Msun year$^{-1}$)$^{-1}$ for stripped stars, which is about an order of magnitude higher than what massive stars produce. The yields of ionizing photons are affected by metallicity (for details, see \appref{app:model_variations}).

%\YG{Update the \Iion files before publishing (no big difference expected, just to be exact).}

% % % % %   Z-z relation   % % % % % 

\paragraph{Relation between redshift and metallicity} ~\\
%
% Metallicity affects the yields of ionizing photons
%   - differences in our models, refer to appendix for details
\noindent The emission rate of \HI-ionizing photons from massive main-sequence stars is known to increase with decreasing metallicity \citep[e.g.,][]{2015ApJ...800...97T}. The emission from stripped stars is only mildly affected by metallicity, leading to about a factor of two higher emission rates of \HI-ionizing photons at $Z = 0.002$ compared to at solar metallicity (\citetalias{2017A&A...608A..11G}; \citetalias{2018A&A...615A..78G}). The yields of \HeII-ionizing photons show large fluctuations with metallicity for single star populations and large variations between different spectral synthesis models, but remain relatively constant for stripped stars with $Z \geq 0.002$ (see \appref{app:model_variations}).

% Metallicity changes over cosmic time -- yields change over cosmic time
The early stellar populations that reionized the Universe were likely metal-poor \citep[e.g.,][]{2007A&A...462..429B, 2008MNRAS.383.1439G}.
% Model for how the yields change over cosmic time: z-Z relation
%   - first, just simple approach
%   - next step: upgrade to complex functions of Langer & Norman (2006)
%   - provide a figure in the appendix?
%$Z/Z_{\odot} = 10^{-0.15z}$ \citep{2007iuse.book..435K}
We account for the change of metallicity over redshift by assuming that all stellar populations have the average metallicity, which evolves with redshift following the relation $\log_{10} \langle Z/Z_{\odot} \rangle = 0.153 - 0.074z^{1.34}$ of \citet[][see also \citealt{2007iuse.book..435K}]{2017ApJ...840...39M}, where we assume solar metallicity to be $Z_{\odot} \equiv 0.014$ \citep{2009ARA&A..47..481A}. We then interpolate the yields of ionizing photons over metallicity using log-scales for both the metallicity and the yields to reach a trend as smooth as possible. Next, we use the scaling between metallicity and redshift to obtain the appropriate yields of ionizing photons for each redshift. Our method is approximate since a spread of metallicity have been observed at each redshift \citep[see][and references therein]{2014ARA&A..52..415M}. However, the above mentioned weak sensitivity of the \HI-ionizing radiation to metallicity allows us to consider only a mean metallicity per redshift.
%However, the emission rates of \HI-ionizing photons from stripped stars are only weakly sensitive to metallicity variations (variations within $\sim 50\%$, \citetalias{2018A&A...615A..78G}). 
The already uncertain predictions for the \HeII-ionizing emission are significantly affected by metallicity variations.

% % % % %   Reaching the production rates of ionizing photons   % % % % %

\paragraph{Production rates of ionizing photons, \ndotion}~\\
%
% We can create the rate with which the stars produce ionizing photons over cosmic time by multiplying yields with cosmic SFR
%   - give the formula: first for ndotion0 and 2?, then just refer to the psi of MD14 and their equation X 
\noindent We describe the rate with which stellar populations produce ionizing photons over cosmic time, \ndotion, by multiplying the yields of ionizing photons, \Iion, with the cosmic star-formation rate density, $\psi (z)$,
\begin{equation}\label{eq:ndotion_stars}
\begin{aligned}
&\ndotionz (z) = \Iionz (z)\, \psi (z) \\
&\ndotiont (z) = \Iiont (z) \, \psi (z) .
\end{aligned}
\end{equation}
We employ the cosmic star-formation history of \citet[][their Eq.~15]{2014ARA&A..52..415M} together with the yields of ionizing photons that we presented previously. %, also consistent with \citealt{2015ApJ...810...71F}
% Uncertainties
%   - different cosmic SFH to compare with?
%   - already discussed uncertainties of Iion in the two previous subsections
The star-formation rate density of \citet{2014ARA&A..52..415M} reaches a peak at about $z\sim 2-3$ and decreases for both higher and lower redshifts. Star-formation in galaxies at high redshift that are too faint to be observed has been suggested to provide a large fraction of the ionizing emission and impact the reionization \citep[e.g.,][]{2012ApJ...752L...5B, 2014MNRAS.442.2560W}. However, the relative importance of stripped stars to massive stars remains the same independent on the star-formation history. Therefore, we do not expect significant changes in the relative contribution from stripped stars if a different star-formation history is considered, although the estimated time of reionization may change.
%Here, we take the conservative approach of \citet{2014ARA&A..52..415M} since our aim is to quantify the effect of stripped stars, which we expect to be present at all redshifts.
% \citep[\YG{REF}][for the current measurements see \citealt{2012ApJ...754...83B, 2012ApJ...752L...5B}]{}. 
%Here, we assume the star-formation rate of \citet{2014ARA&A..52..415M}, but discuss more in \secref{sec:reionization} the effect of a shallower star-formation history at high redshifts. 

% Assumption of the formula -- instantaneous emission of ionizing photons
%   - why it is ok
%   - what would be a better method
\eqref{eq:ndotion_stars} assumes that all ionizing photons that stellar populations produce are emitted at the instant that the stars are formed. In reality, stars emit ionizing photons during the entirety or parts of their lifetimes, as shown in \figref{fig:Qt}. However, approximating the ionizing emission to be instantaneous is justified when a time interval that is much longer than the stellar lifetimes is considered. This is the case for cosmic reionization, which lasts for several hundred Myr and most of the emission of ionizing photons from stellar populations extends over about 50~Myr (\citetalias{2019A&A...629A.134G} and \figref{fig:Qt}).

% % % % %   Assuming an escape fraction   % % % % %

\paragraph{Escape fraction, \fesc} ~\\
\begin{figure}
\centering
\includegraphics[width=.8\hsize]{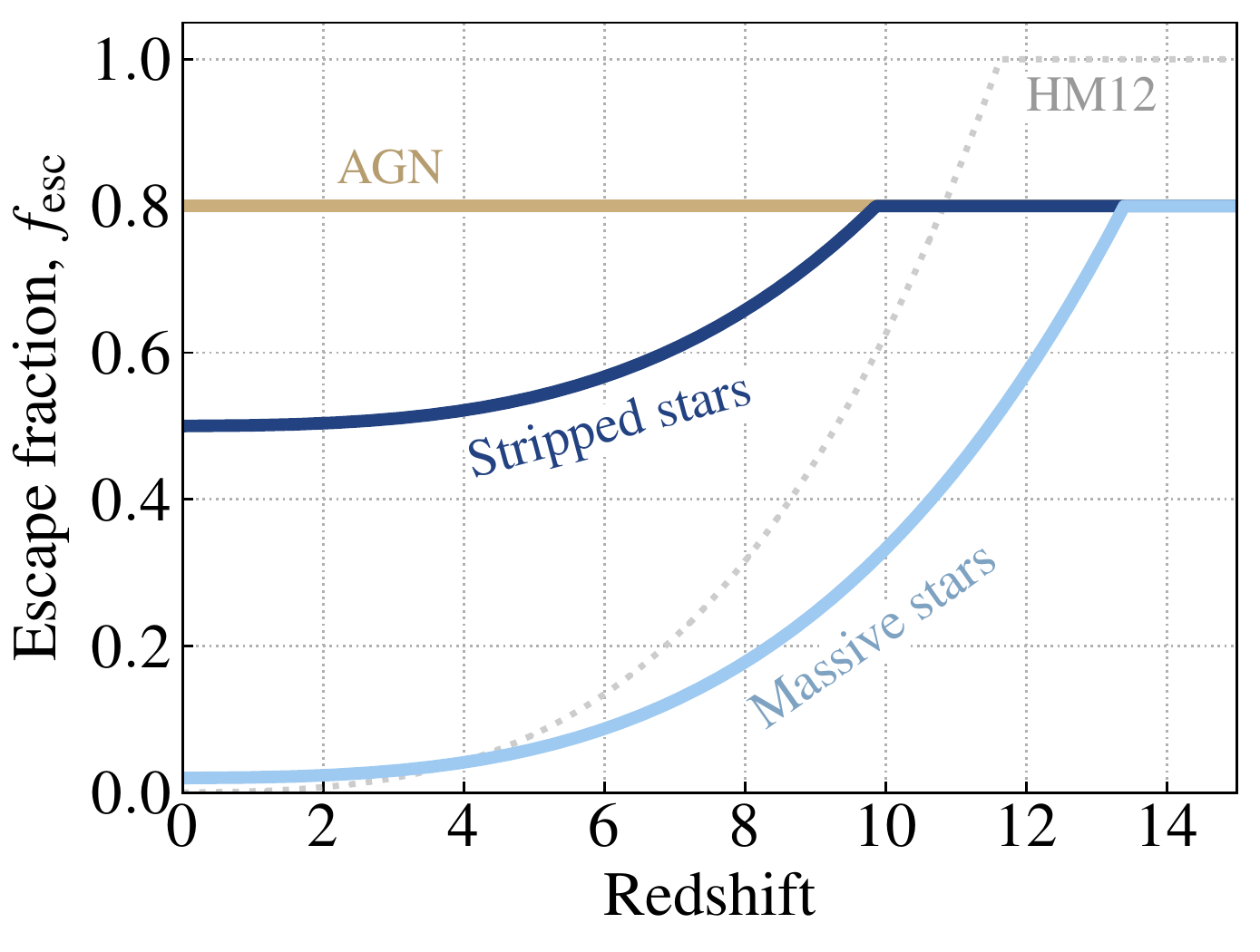}
\caption{The redshift dependent escape fractions that we assume in our standard model for stripped stars (dark blue), massive stars (light blue), and AGN (beige). The functions for the escape fraction for stripped stars and massive stars are given in \eqreftwo{eq:fescs}{eq:fescm}, respectively. We also show the escape fraction of \citet{2012ApJ...746..125H} for stellar populations as a gray dotted line and marked as HM12.}
\label{fig:fesc}
\end{figure}
\noindent We follow the approach by \citet{2012ApJ...746..125H} when simulating the escape fraction of ionizing photons over cosmic time. Their function for the escape fraction increases from 0\% in the local Universe up to 20\% at $z \sim 7$ and stalls at 100\% at $z \sim 11$. We modify the function slightly to match what we consider realistic for massive stars and for stripped stars separately. We shift the escape fraction up to 2\% for massive stars and 50\% for stripped stars in the local Universe. We choose the lower limit for massive stars because star-forming regions in the nearby Universe have been observed to have an escape fraction of a few percent \citep[e.g.,][]{2013ApJ...779...65M, 2013A&A...558A.134D, 2016ApJ...823...64L, 2016Natur.529..178I, 2016MNRAS.461.3683I, 2018ApJ...869..123S, 2019MNRAS.483.5380T}, which simulations agree well with \citep{2011A&A...530A..87P, 2012ApJ...747..100S, 2015MNRAS.451.2544P, 2015MNRAS.451.1939R, 2015MNRAS.453..960M, 2016ApJ...819...81R}. However, simulations also show that the escape fraction increases with time after star-formation has stopped, and it is likely to reach $\sim$50\% after a few tens of Myr because of the strong stellar winds and supernovae from massive stars \citep{2014ApJ...788..121K, 2017MNRAS.470..224T, 2019MNRAS.486.2215K, 2019arXiv190807549K}. At this time, stripped stars are present in the stellar populations and many are yet to be created \citepalias[see][]{2019A&A...629A.134G}. %Recent observations also show that the escape fraction varies depending on the viewing angle \citep{2019arXiv190408186R}, hinting at a clumpy interstellar environment with possible channels of 
Studies presenting compilations over a range of redshifts indicate that the escape fraction was higher at early times \citep[][see however \citealt{2008ApJ...672..765G}]{2006MNRAS.371L...1I, 2016ApJ...829...99F, 2018arXiv180601741F}. Because galaxies were small in the early Universe \citep{2017arXiv171102090B}, they had a larger relative surface area, which leads to a higher leakage of ionizing photons \citep{2009ApJ...693..984W, 2014MNRAS.442.2560W}. This motivates the increase of the function for the escape fraction with higher redshift. 
To avoid completely transparent galaxies, we also truncate the functions for the escape fractions so that they do not reach above 80\%. The function of \citet{2012ApJ...746..125H} is relatively steep and we therefore also multiply the function with a factor of 0.5 to allow for lower escape fractions at higher redshits. The result is the following functions:
\begin{equation}\label{eq:fescs}
\fescs (z) =  0.5 +  0.067 \left( \dfrac{1+z}{7}\right)^{3.4}, \quad \fescs \leq 0.8  .
\end{equation}
\begin{equation}\label{eq:fescm}
\fescm (z) = 0.02 + 0.067 \left( \dfrac{1+z}{7}\right)^{3.4}, \quad \fescm \leq 0.8
\end{equation}
\citet{2012ApJ...746..125H} do not allow \HeII-ionizing photons produced in galaxies to emerge into the IGM. We allow the emission of \HeII-ionizing photons, using the same escape fractions with the motivation that photons may be able to escape through open holes in the interstellar medium, which means that no material can stop the radiation from reaching the IGM. Recent observational studies high redshift galaxies support this idea \citep[see][]{2019arXiv190800556J, 2019arXiv190408186R}. We show our assumed escape fractions as a function of redshift in \figref{fig:fesc}.

The escape fraction is an uncertain parameter that affects the resulting time of reionization and the contribution from the different sources of ionizing radiation. To explore the dependence on the escape fraction, we also consider three models with constant escape fractions, one with high escape fractions, one with low escape fractions, and one in which stripped stars have a high escape fraction and massive stars have a low escape fraction (see \tabreftwo{tab:models}{tab:fesc}).    %We discuss the results from these models in more detail in \appref{app:fesc}.

% _______ _ _ _ ________ GO TO AGN _________ _ _ _ _________

% The ionizing emission from quasars
\subsubsection{Active galactic nuclei}\label{sec:AGN_ndotion}

% Background: 
%   - AGN are accreting black holes and the accretion disk emits radiation similar to BBs for each ring and with a different temperature
%   - spectral shape of AGN
%   - occurrence over cosmic time
The accretion discs around supermassive black holes in the center of galaxies are so hot that they radiate both hydrogen and helium ionizing photons. Their emission is likely harder than that from stellar populations because parts of the accretion discs can reach higher temperatures than stars. The ionizing spectra of AGN are usually approximated by a power-law, $L_{\nu} \propto \nu ^{\alpha}$, for which the slope varies for individual objects ($-3 \lesssim \alpha \lesssim -1$) but can be averaged to $\alpha = -1.7$ \citep[][see also \citealt{2002ApJ...565..773T}]{2015MNRAS.449.4204L}.
%\figref{fig:SED_sources} compares the spectral hardness of AGN with that of stellar populations and demonstrates that AGN are more efficient in producing \HeII-ionizing photons compared to stellar populations. However, w
We note that the spectra of AGN are unexplored observationally for the wavelength range short-wards of $\sim 350$~\AA\ and for photon energies smaller than $\sim 0.1$~keV \citep[e.g.,][]{2018ApJ...869..159U}. In this regime, the spectral slope is typically extrapolated from longer wavelengths \citep[e.g.,][]{2012ApJ...746..125H}. 
The number density of AGN has been observed to reach a peak at $z\sim 2-3$ and to have a steep decline for both higher and lower redshifts \citep[e.g.,][]{2007ApJ...654..731H}.

% Emissivity of AGN from Haardt & Madau (2012)
%   - describe the origin on the formula (Hopkins+07) 
%   - caveats of the model 
%      o doesn't cover more than z~6?
%      o few datapoints at high z, could be steeper (MH15?)
Calculating the ionizing emissivity of AGN, $\ndotionz$ and $\ndotiont$, requires combining the number density of AGN and their spectra. %The ionizing emissivity, which has units of erg~s$^{-1}$~Hz$^{-1}$~Mpc$^{-3}$, relates to the production rate of ionizing photons: \begin{equation}\label{eq:ndotion_AGN} \ndotionz = \dfrac{{\epsilon}_{912}(z)}{(1+z)^{3}} \times \dfrac{\nu}{E} = \dfrac{{\epsilon}_{912}(z)}{(1+z)^{3}} / h.\end{equation} In \eqref{eq:ndotion_AGN}, $E$ is the photon energy and $h$ is the Planck's constant. 
We employ the ionizing emissivity of \citet[][see their Eq.~25]{2019MNRAS.485...47P} to represent the production rate of \HI-ionizing photons from AGN over cosmic time. 
%  The spectral shape of AGN - observations and extrapolations
%   - how we create eps_2 by assuming a spectral shape.
For the production rate of \HeII-ionizing photons from AGN, we use the \HI\ ionizing emissivity and extrapolate to higher frequencies using the average spectral shape of AGN from \citet{2015MNRAS.449.4204L}. 
%We note that the production rate of \HeII-ionizing photons from AGN is predicted using an extrapolated part of the spectra and thus uncertain.

% Assumptions for escape fraction and discuss whether it is too optimistic? put 70%?
The escape fraction from AGN is commonly assumed to be unity because of their violent outflows and strong ionizing emission. However, recent studies suggest that the escape fraction might be somewhat smaller \citep{2016MNRAS.462.2478C, 2017MNRAS.465..302M}. We follow the findings of \citet{2016MNRAS.462.2478C} and adopt an escape fraction of 80\% for AGN (which is also the cap for our stellar populations), assuming that it is constant over cosmic time.

% Describes the analytical method to solve a simple model for reionization
% Discussing both reionization of H and He
% Introduces the quantities inside the differential equation
% Explains what co-moving means in a very subtle way so that the reader does not feel stupid
\subsection{Cosmic reionization: a simple semi-analytic model}\label{sec:analytical_model}

% Goal to reach: estimate impact of stripped stars on the epoch of reionization for H and He. 
% Method: analytical, seen in Haardt & Madau 2012, also other (citations) -- give equation (same method for both H and He)
We estimate the contribution from stripped stars to the reionization of hydrogen and helium by calculating their impact on the volume filling factor of ionized gas, $x$, over cosmic time. The volume filling factor is a dimensionless parameter that describes the fraction of gas in the IGM that is ionized in terms of volume. We follow a simple approach, described for example in \citet{1999ApJ...514..648M} and \citet{2012ApJ...746..125H}, where the rate of change of the volume filling factor of ionized gas is described as the balance between the rate with which ionizing photons reach the atoms or ions in the IGM and the rate with which the ions recombine. The rate of change of the volume filling factors of ionized hydrogen and helium can be written in the form of the following differential equations:
\begin{equation}\label{eq:vol_fill}
\begin{aligned}
&\dfrac{dx_{\HII}}{dt} = \dfrac{f_{\text{esc}}\ndotionz}{\langle n_{\text{H}} \rangle} - \dfrac{x_{\HII}}{\langle t_{\text{rec, \HII}} \rangle} \\
&\dfrac{dx_{\HeIII}}{dt} = \dfrac{f_{\text{esc}}\ndotiont}{\langle n_{\text{He}} \rangle} - \dfrac{x_{\HeIII}}{\langle t_{\text{rec, \HeIII}} \rangle},
\end{aligned}
\end{equation}
where $\langle n \rangle$ is the mean number density of a considered element in the IGM, and $\langle t_{\text{rec}} \rangle$ is the mean recombination timescale for a given ion. The combination of the escape fraction and the production rate of ionizing photons, $f_{\text{esc}}\ndotion$, describes the rate at which ionizing photons reach the IGM. We use the subscripts H and He to denote the quantities for hydrogen and helium, respectively. The escape fraction is assumed to be the same for both hydrogen and helium since we assume it primarily is an effect of geometry within the host galaxy \citep[cf.][]{2019arXiv190800556J, 2019arXiv190408186R}. However, we discuss the possibility and impact of a wavelength dependent escape fractions in \secref{sec:SED_results}.
We use quantities for a co-moving cosmological volume throughout this paper, unless other is stated. %In this way, we follow the evolution of a region that expands with the expansion of the Universe over cosmic time. 

% Recombination timescale and geometrical effects and density
%   - density of IGM & primordial composition
%   - how to calculate the recombination timescale (brief) 
%   - assumption that HI and HeI become reionized together and then later comes HeII. This means that the chi is slightly smaller for HI and HeI 
%We assume that the IGM has a number density of hydrogen of $\langle n_{\text{H}} \rangle = 1.9 \times 10^{-7} \; (1+z)^3$~cm$^{-3}$ \citep[][]{2012ApJ...746..125H} and that the composition in the IGM is primordial \citep[e.g.,][]{1967ApJ...148....3W}, i.e., for every helium atom there are approximately twelve hydrogen atoms. The electron density 
We calculate the mean recombination time of ionized hydrogen and helium following \citet{2012ApJ...746..125H}:
\begin{equation}\label{eq:trec}
\begin{aligned}
%&\langle t_{\text{rec, }\HII} \rangle = [\chi_{\HII}\; \langle n_{\text{H}} \rangle\; \alpha_{B, \HII} \; C_{\text{IGM}}]^{-1} \\
%%&\langle t_{\text{rec, }\HeII} \rangle = [\chi_{\HeII} \langle n_{\text{He}} \rangle \alpha_{B, \HeII} C_{\text{IGM}}]^{-1} \\
%&\langle t_{\text{rec, }\HeIII} \rangle = [\chi_{\HeIII}\; \langle n_{\text{He}} \rangle \; \alpha_{B, \HeIII} \; C_{\text{IGM}}]^{-1}.
&\langle t_{\text{rec, }\HII} \rangle = [n_e\; \alpha_{B, \HII} \; C_{\text{IGM}}]^{-1} \\
%&\langle t_{\text{rec, }\HeII} \rangle = [\chi_{\HeII} \langle n_{\text{He}} \rangle \alpha_{B, \HeII} C_{\text{IGM}}]^{-1} \\
&\langle t_{\text{rec, }\HeIII} \rangle = [n_e \; \alpha_{B, \HeIII} \; C_{\text{IGM}}]^{-1},
\end{aligned}
\end{equation}
In \eqref{eq:trec}, the clumpy structure of the hydrogen and helium in the IGM is approximated by a redshift dependent clumping factor, $C_{\text{IGM}} = 1 + 43z^{-1.71}$ \citep{2012ApJ...746..125H}. It is likely that the hydrogen and singly-ionized helium gas are somewhat differently distributed during their respective reionizations, but for simplicity we use the same clumping factor. However, the choice of clumping factor does not significantly affect our results. We assume that the IGM has a temperature of $\sim 20\,000$~K during reionization \citep[see e.g.,][]{1994MNRAS.266..343M, 2019ApJ...874..154D}, meaning that the recombination coefficients are set to $\alpha_{B, \HII} = 1.43 \times 10^{-13}$~cm$^{3}$~s$^{-1}$ and $\alpha_{B, \HeIII} = 9.08 \times 10^{-13}$~cm$^{3}$~s$^{-1}$ \citep[for Case~B type recombination, following][]{2006agna.book.....O}. The electron density, $n_e$, accounts for the free electrons from ionization of both hydrogen and helium. We assume that the reionizations of \HI and \HeI were coupled as the number of \HeI-ionizing photons were sufficient for \HeI reionization to occur prior to \HI reionization, but the mean-free path of the photons were limited by the more abundant hydrogen \citep{1994MNRAS.266..343M}. We then assume that reionization of \HeII occurred after that of \HI and \HeI. Furthermore, we assume that the ionization fronts were thin compared to the size of the ionized bubbles they enclosed, which is a good approximation for all but the very first stages of reionization \citep[for a discussion, see][]{1999ApJ...514..648M}. This means that there were free electrons from ionized hydrogen and singly ionized helium in the ionized bubbles during \HI-reionization, which results in an electron density of $n_e = 1.08 \, \langle n_{\text{H}} \rangle$, where the mean number density of hydrogen is  $\langle n_{\text{H}} \rangle = 1.9 \times 10^{-7} \; (1+z)^3$~cm$^{-3}$ \citep[][]{2012ApJ...746..125H}. We have assumed that the composition in the IGM is primordial \citep[e.g.,][]{1967ApJ...148....3W}, meaning that for every helium atom there are approximately twelve hydrogen atoms. During \HeII-reionization, the electron density was higher inside the ionized bubbles because helium was fully ionized and, therefore, $n_e = 1.17 \, \langle n_{\text{H}} \rangle$. %For the calculation of the recombination times, physical units are used \edit{instead of the otherwise used comoving units}. 
%\edit{For the calculation of the recombination times, we do not use the comoving units.}
The resulting recombination timescale is about 500~Myr at $z \sim 6$ for recombination of hydrogen and about seven times shorter for recombination of \HeIII to \HeII at the same time. 

% translate from redshift to time after birth of Universe
When solving \eqref{eq:vol_fill}, we integrate over time assuming a flat cosmology in the standard $\Lambda$CDM theory and apply $H(z) = H_0 (\Omega_M (1+z)^3  + \Omega_{\Lambda})^{1/2}$. For the cosmological parameters, we use the latest results from the Planck satellite \citep[$\Omega_M = 0.31$, $\Omega_{\Lambda} = 0.69$, and $H_0 = 68$~km~s$^{-1}$~Mpc$^{-1}$,][]{2016A&A...594A..13P}. %The age of the Universe, $t_{\text{age}}$, is then related to the cosmic redshift, $z$, by the following integration:
%\begin{equation}\label{eq:t_age}
%t_{\text{age}} = \int _z^{\infty} \dfrac{dz'}{H(z')(1+z')} .
%\end{equation}

% Table summarizing the two models we consider for the ionizing emission over cosmic time
\begin{table}
\begin{center}
\caption{The four models that we consider for our calculation of cosmic reionization. They differ from each other by the assumed escape fraction for the considered ionizing sources.}
\label{tab:models}
\begin{tabular}{lccc}
\toprule\midrule
Model & \multicolumn{3}{c}{Escape fraction (\fesc)} \\
& Stripped stars & Massive stars & AGN \\
\midrule
Standard & \eqref{eq:fescs} & \eqref{eq:fescm} & 80\% \\
High-\fesc & 80\% & 20\% & 80\% \\
Low-\fesc & 30\% & 5\% & 80\% \\
Combination & 80\% & 5\% & 80\% \\
\bottomrule
\end{tabular}
\end{center}
{\tablefoot{\footnotesize We use the models presented in \citetalias{2019A&A...629A.134G} to represent the ionizing emission from stripped stars, the predictions from \code{Starburst99} for the contribution from massive main-sequence and WR stars \citep{1999ApJS..123....3L, 2010ApJS..189..309L}, and the emissivity presented in \citet{2019MNRAS.485...47P} to represent the ionizing emission from AGN. For details, see \secref{sec:ionizing_sources}.\par}}
\end{table}

% Where do we get the different quantities from? 
In \secref{sec:ionizing_sources}, we described how we estimate the production rates of ionizing photons and the escape fraction of ionizing photons for the various ionizing sources that we consider. We use these together with the calculated recombination times to numerically solve \eqref{eq:vol_fill} and present the solutions in \secref{sec:reionization}. 
% We create several models, and treat in particular the standard model here, the others in the appendix
We create four models for the cosmic emissivity of ionizing photons distinguished by different assumptions for the escape fraction of ionizing photons, as summarized in \tabref{tab:models}. The standard model uses the functions for the escape fraction as described by \eqreftwo{eq:fescs}{eq:fescm}. Since the escape fraction is uncertain, we also consider a high-\fesc and a low-\fesc model, which both assume constant escape fractions such that the reality likely resides somewhere between the two models. For the high-\fesc model, we assume an escape fraction of 20\% for massive stars and 80\% for stripped stars, while for the low-\fesc model, we instead assume an escape fraction of 5\% for massive stars and 30\% for stripped stars. We also create a combination model in which we assume that the escape fraction is 5\% for massive stars and 80\% for stripped stars. For all models we assume the constant escape fraction of 80\% for AGN. %We focus on the results from our standard model, but also discuss the results from the optimistic and pessimistic models. 

%________________________________________________________________

\section{Ionizing emissivity over cosmic time}\label{sec:SED_results}

% FIGURE of the fesc*ndotion0 and fesc*ndotion2 with cosmic time
\begin{figure*}
\centering
\includegraphics[width=0.32\textwidth]{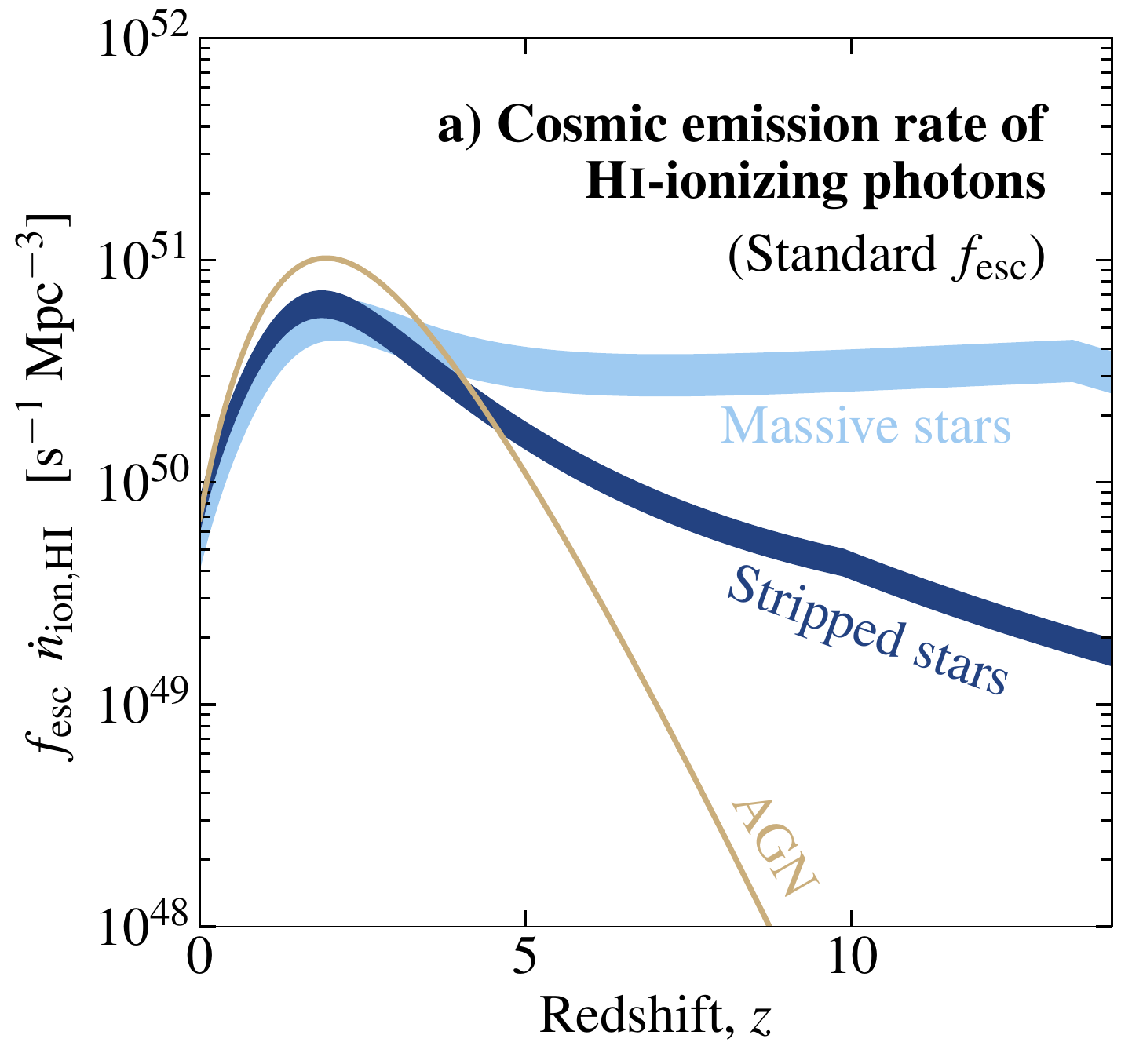}
\includegraphics[width=0.32\textwidth]{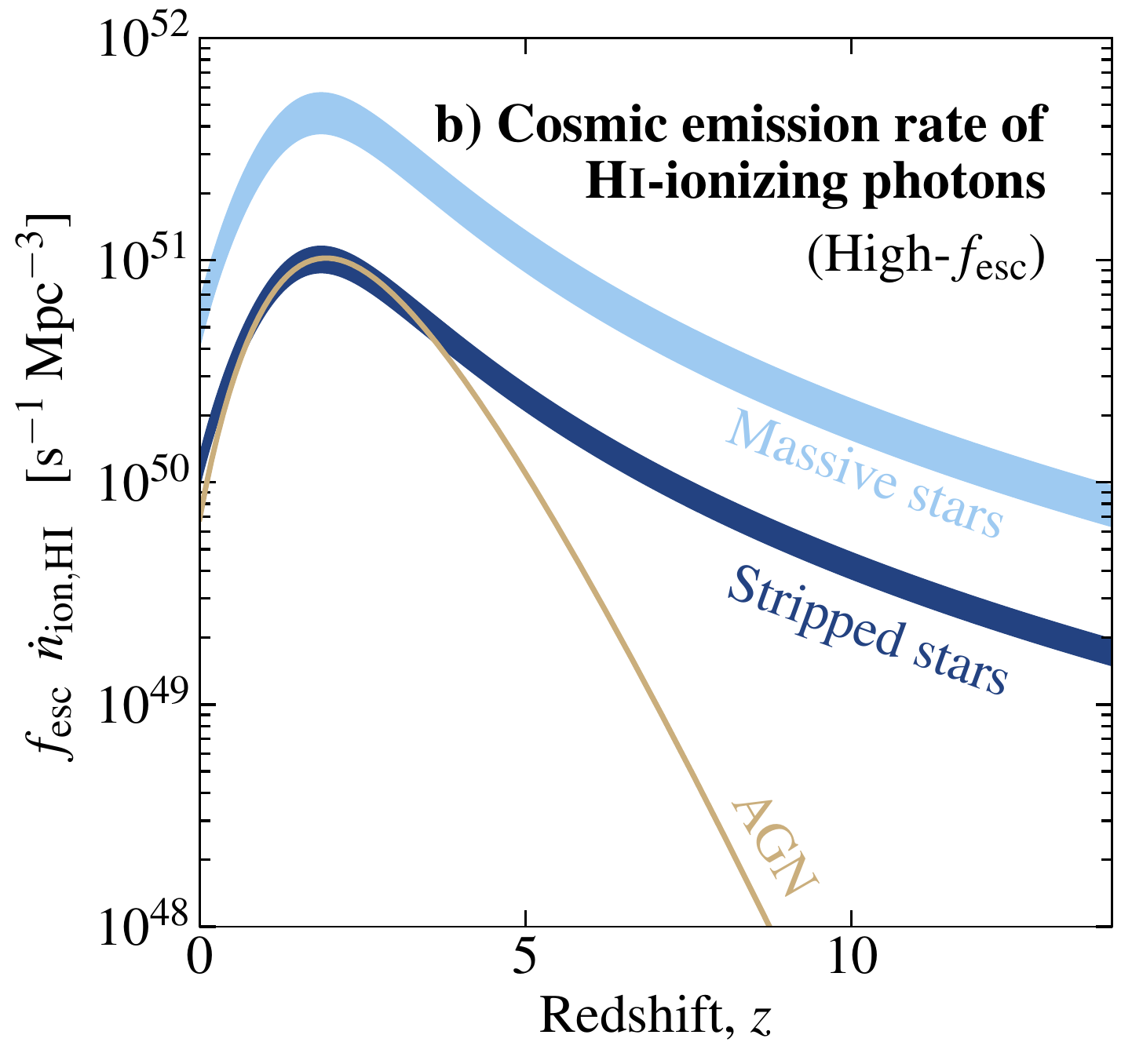}
\includegraphics[width=0.32\textwidth]{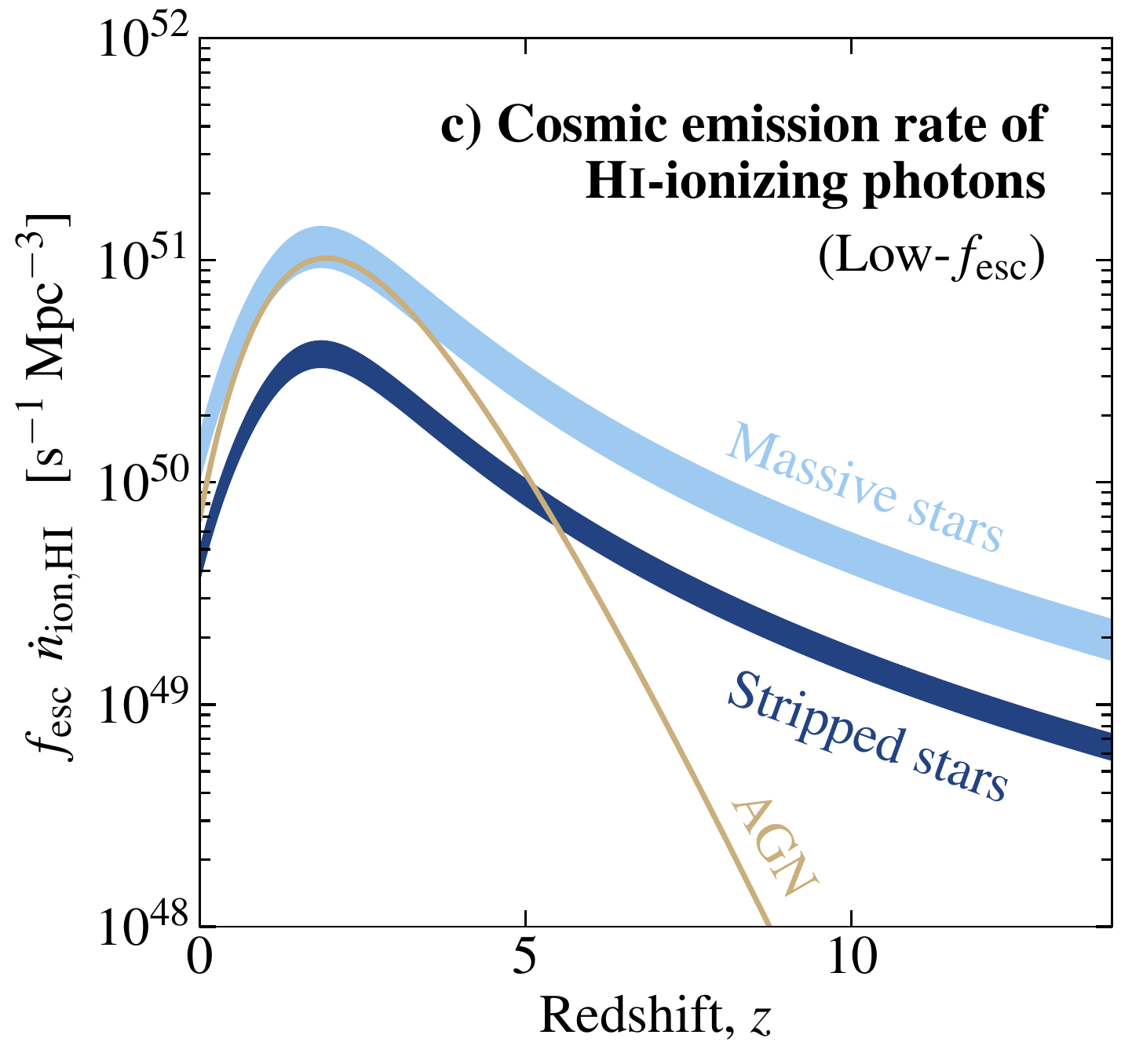}
\includegraphics[width=0.32\textwidth]{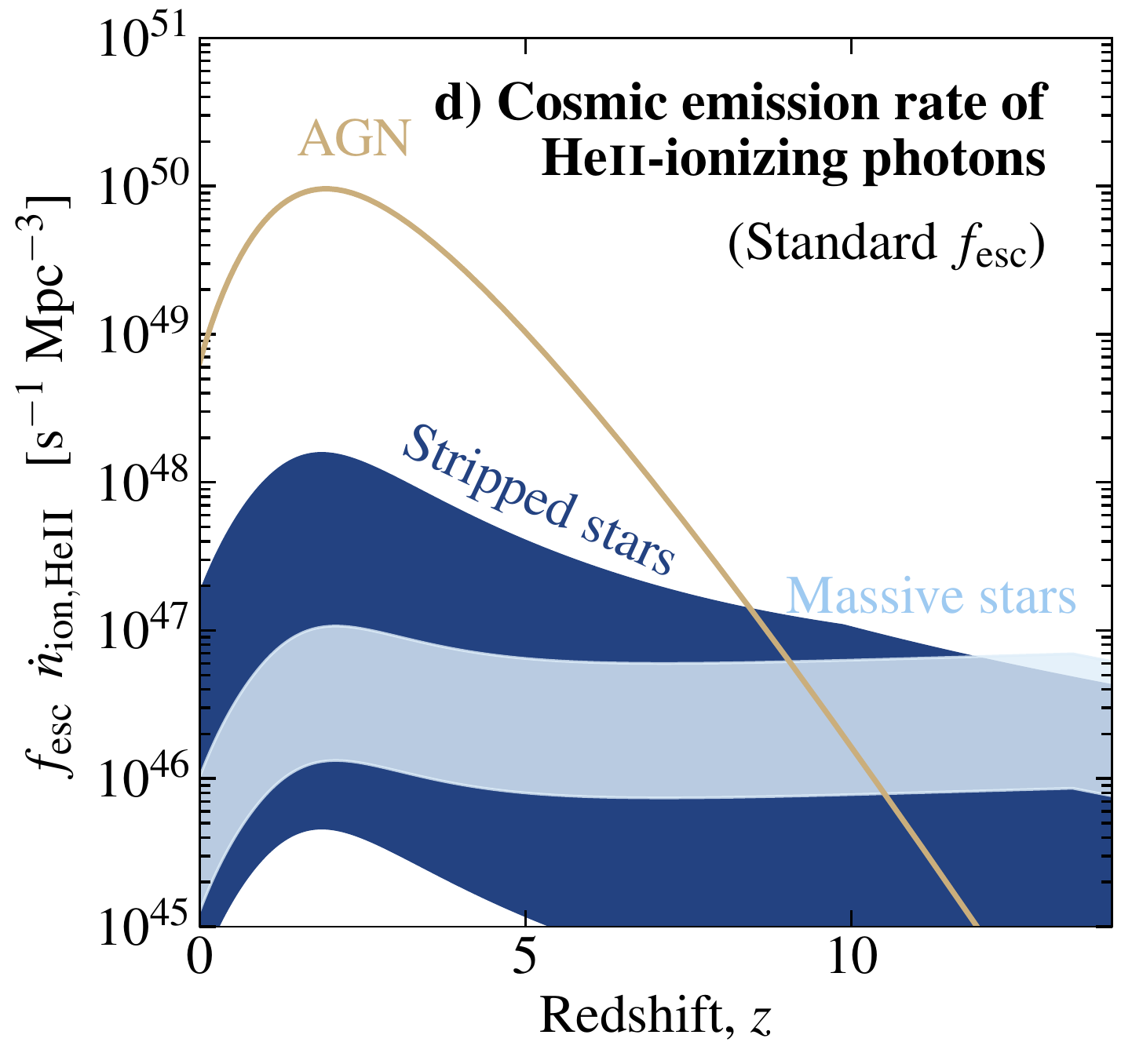}
\includegraphics[width=0.32\textwidth]{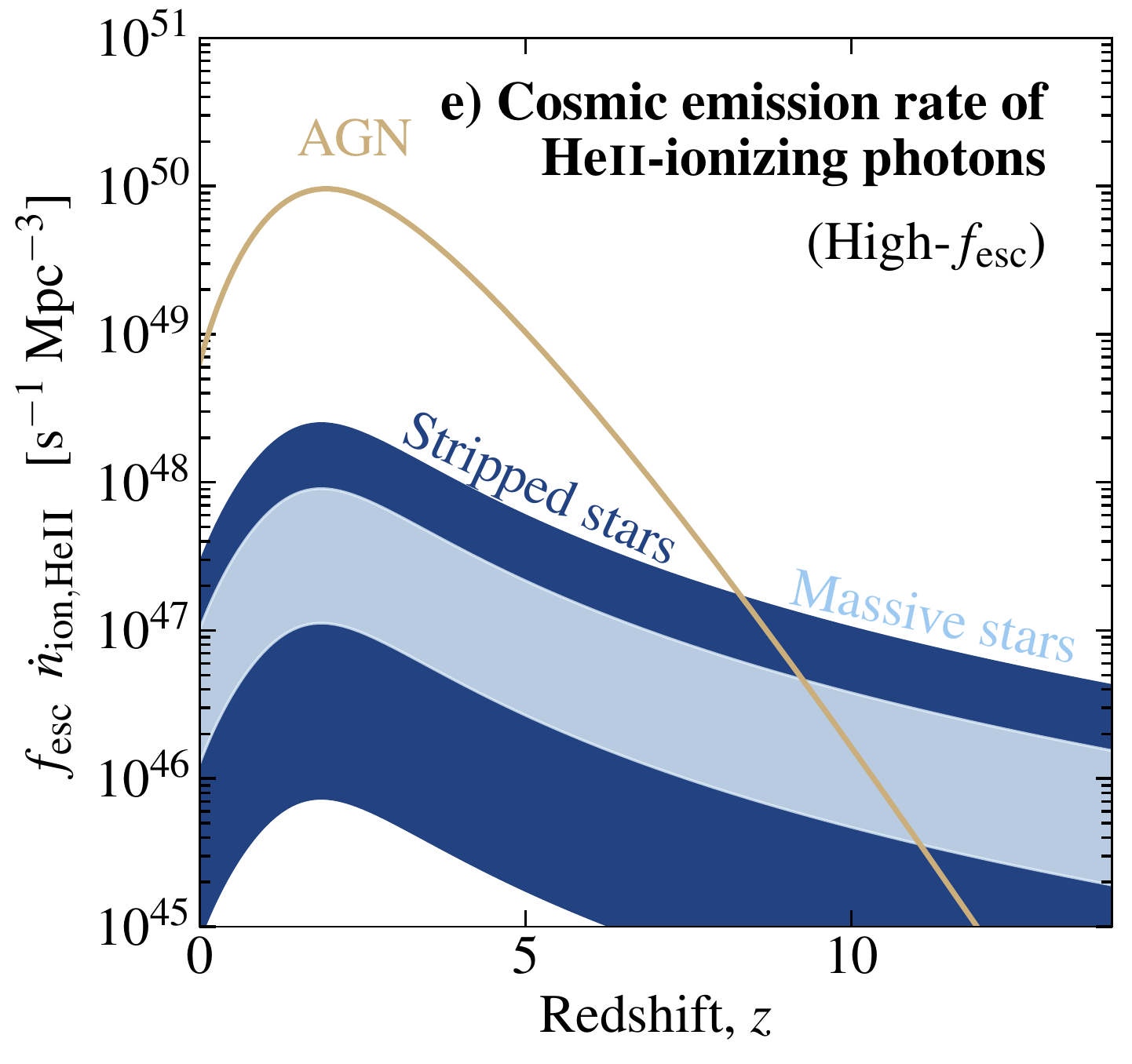}
\includegraphics[width=0.32\textwidth]{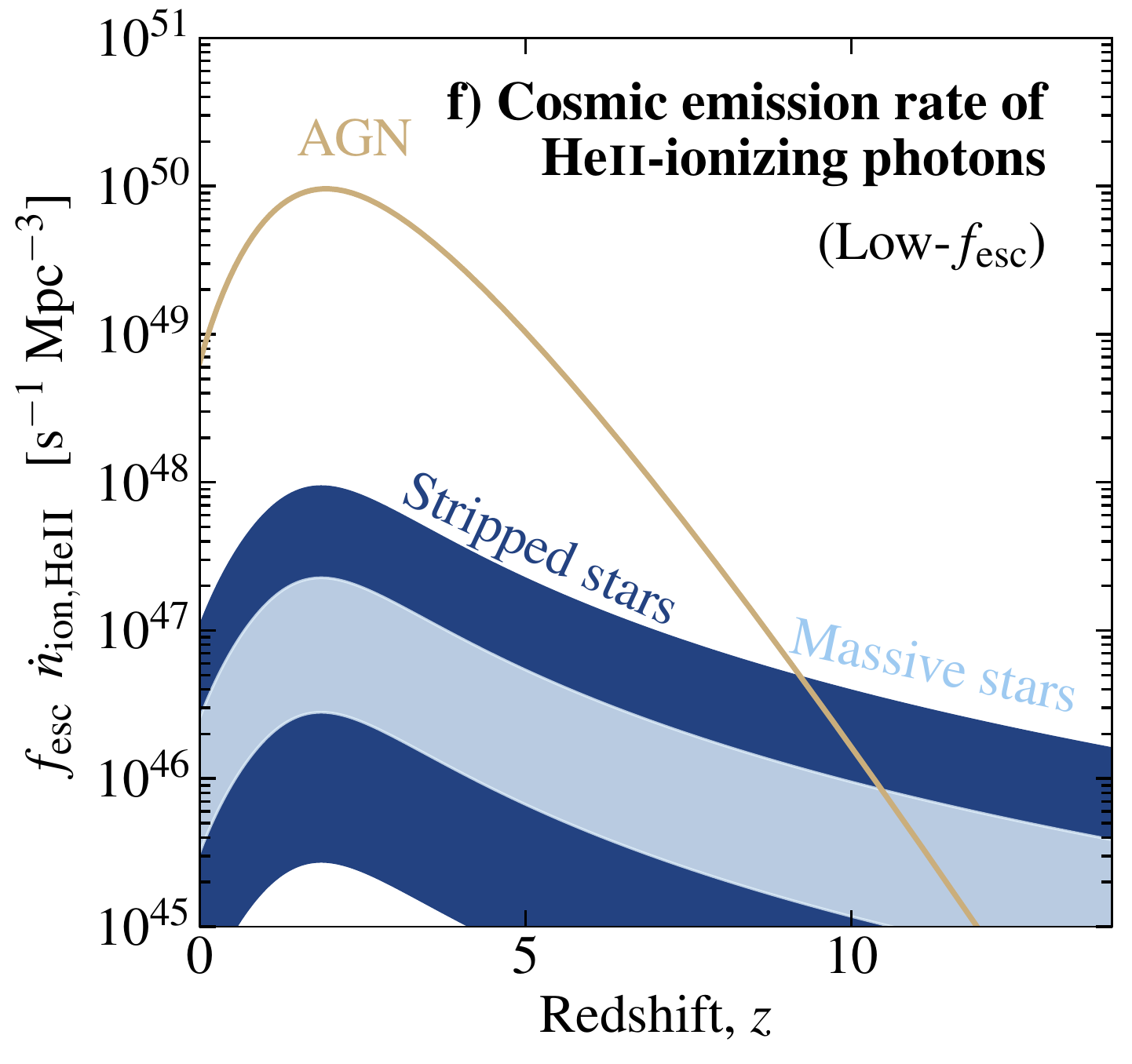}
\caption{The rate with which ionizing photons reach the intergalactic medium. Panel a, b, and c show the rate for \HI-ionizing photons and panel d, e, and f for \HeII-ionizing photons. The left column shows the results for our standard model, the middle column for our high-\fesc model, and the right column for our low-\fesc model (see \tabref{tab:models}). We show the contribution from stripped stars in dark blue, from massive stars in light blue, and from AGN in brown. The extent of the colored regions represent the spread that metallicity variations give rise to. The results for the combination model (see \tabref{tab:models}) are the same for the stripped stars as in the high-\fesc and for the massive stars as in the low-\fesc model.}
\label{fig:ndotion}
\end{figure*}

% Emissivity of ionizing photons over cosmic time
\subsection{Cosmic emission rates of \HI- and \HeII-ionizing photons}

%   - discuss the relative contributions 
% How do we compute the emission rates?
%Using \eqref{eq:ndotion_stars} and Eq.~25 from \citet{2019MNRAS.485...47P}, we calculate the production rates of \HI- and \HeII-ionizing photons from stellar populations and AGN. To obtain the emission rate of ionizing photons that reach the IGM, we multiply the production rates of ionizing photons with the escape fraction for each considered source. 
In \figref{fig:ndotion}, we show the rates at which ionizing photons reach the IGM for the different models described in \tabref{tab:models} and as a function of redshift. These rates are calculated by multiplying the escape fraction with the production rates of ionizing photons (\eqref{eq:ndotion_stars} and Eq.~25 from \citealt{2019MNRAS.485...47P} for stars and AGN, respectively). 

%  HI-ionizing photons
The figure shows that stripped stars potentially played important roles as sources of both \HI- and \HeII-ionizing photons, but during different epochs in cosmic time. \figref{fig:ndotion}a shows the emission rate of \HI-ionizing photons from stars and AGN as a function of redshift for our standard model. The model predicts that stripped stars were responsible for a few up to about $25\%$ of the total emission rate of \HI-ionizing photons prior to $z = 6$. Massive stars are predicted to have been the main provider of \HI-ionizing photons at early times and AGN became important for $z < 6$. Around $z = 2-3$, AGN provided the majority of the \HI-ionizing photons and, therefore, likely played a major role in keeping the intergalactic hydrogen ionized. 
%Because we show the results for our standard model here that assumes redshift dependent escape fractions, the emission of \HI-ionizing photons from stripped stars exceeds that of massive stars for  $z < 4$. The effect could be exaggerated, however, we note that the ionizing photons that were relevant for \HI-reionization were emitted prior to $z \sim 6$. 
Panels b and c of \figref{fig:ndotion} show the cosmic emission rates of \HI-ionizing photons in our high-\fesc and low-\fesc models respectively. The panels a-c show that the contribution from stripped stars to the cosmic emission rate of \HI-ionizing photons is relatively large in all our models. We show the range of emission rates that metallicity gives rise to as colored bands for the massive stars and stripped stars in \figref{fig:ndotion}, although in our models we employ a mean metallicity at each redshift. The figure shows that the emission rate of \HI-ionizing photons is only weakly affected by the choice of metallicity, suggesting that accounting for a spread of metallicities at each redshift might have little effect on the result. However, we note that a difference of a factor of two in the emission rate of ionizing photons can have important consequences for the predictions for when reionization occurred and which sources provided the ionizing photons. Our models predict that stripped stars produce \HI-ionizing photons at the highest rate when the metallicity is $Z = 0.002$, when their emission rate of \HI-ionzing photons is about 35\% higher than at solar metallicity (see \tabref{tab:Iion}). 

% Comment on how the flatter emission rate matches better with photoionization rate estimates 
The photoionization rate of the IGM has been observed to be remarkably constant over cosmic time \citep[e.g.,][\edit{see also \citealt{2019MNRAS.489.2669M}}]{2013MNRAS.436.1023B}. Since the photoionization rate of the IGM is proportional to the cosmic emission rate of ionizing photons (times the mean free path of ionizing photons -- which is also increasing with decreasing redshift at $z\sim 5$), it appears as if the flatter emission rates presented in our standard model are favored (\figref{fig:ndotion}a), which could suggest that the escape fraction indeed increases with redshift.

% HeII-ionizing photons
\figref{fig:ndotion}d-f show the emission rate of \HeII-ionizing photons from the individual ionizing sources over cosmic time and for our three models. For stars, the \HeII-ionizing photons are produced in the steep Wien part of the spectra, and the emission rate of \HeII-ionizing photons is therefore very sensitive to temperature variations. At lower metallicity, massive stars are hotter and therefore are predicted to emit \HeII-ionizing photons at higher rates.
However, recombination of helium in the stellar atmospheres, which is dependent on the wind mass-loss rates, can easily block large fractions of the emitted \HeII-ionizing flux. Since wind mass-loss rates are uncertain for many types of stars, the emission rate of \HeII-ionizing photons from stellar populations is subject to uncertainties. The large spreads of emission rates of \HeII-ionizing photons from massive stars and stripped stars seen in \figref{fig:ndotion}d-f are therefore due to both the effect of metallicity and uncertainties in stellar atmosphere models \citepalias[cf.][]{2017A&A...608A..11G}. 
We follow the average metallicity-redshift relation of \citet{2017ApJ...840...39M}, but note that, in reality, the importance of stripped stars and massive stars may vary significantly at a given redshift, depending on the metallicity of the individual stellar populations. \figref{fig:ndotion}d-f show that, in all cases, AGN are responsible for most of the emitted \HeII-ionizing photons as they dominate the emission for a large interval in time. However, at very high redshifts, $z \gtrsim 10$, stripped stars and/or massive stars could have dominated the output of \HeII-ionizing photons. 

% Metallicity dependence
%We show the emission rates for a range of metallicities in \figref{fig:ndotion}. For \HeII-ionizing photons, it is clear that metallicity significantly affects the emission rate from stars. The range of emission rates appears small in the case of \HI-ionizing photons. However, we note that a difference of a factor of two can have important consequences for the predictions for when reionization occurred and which sources provided the ionizing photons. 

% Mention uncertainties?

% Show SEDs for the total integrated background spectrum at several redshifts -- mark these in the ndotion diagrams?
\subsection{Spectral energy distribution of the cosmic emissivity}

% What is this?
%   - integrated spectrum hints about which were sources of ionizing radiation since they have (likely) different spectral shapes
%   (- the hardness of the ionizing spectrum may be possible to infer from observables such as the temperature or metal absorption features)
The integrated spectrum of the radiation that emerges into the IGM contains imprints from the sources that produced the ionizing radiation. The hardness of the ionizing radiation can, for example, be used to infer what type of source emitted the radiation, as the different sources are characterized by different spectral shapes. 

\begin{figure*}
\centering
\includegraphics[width=\textwidth]{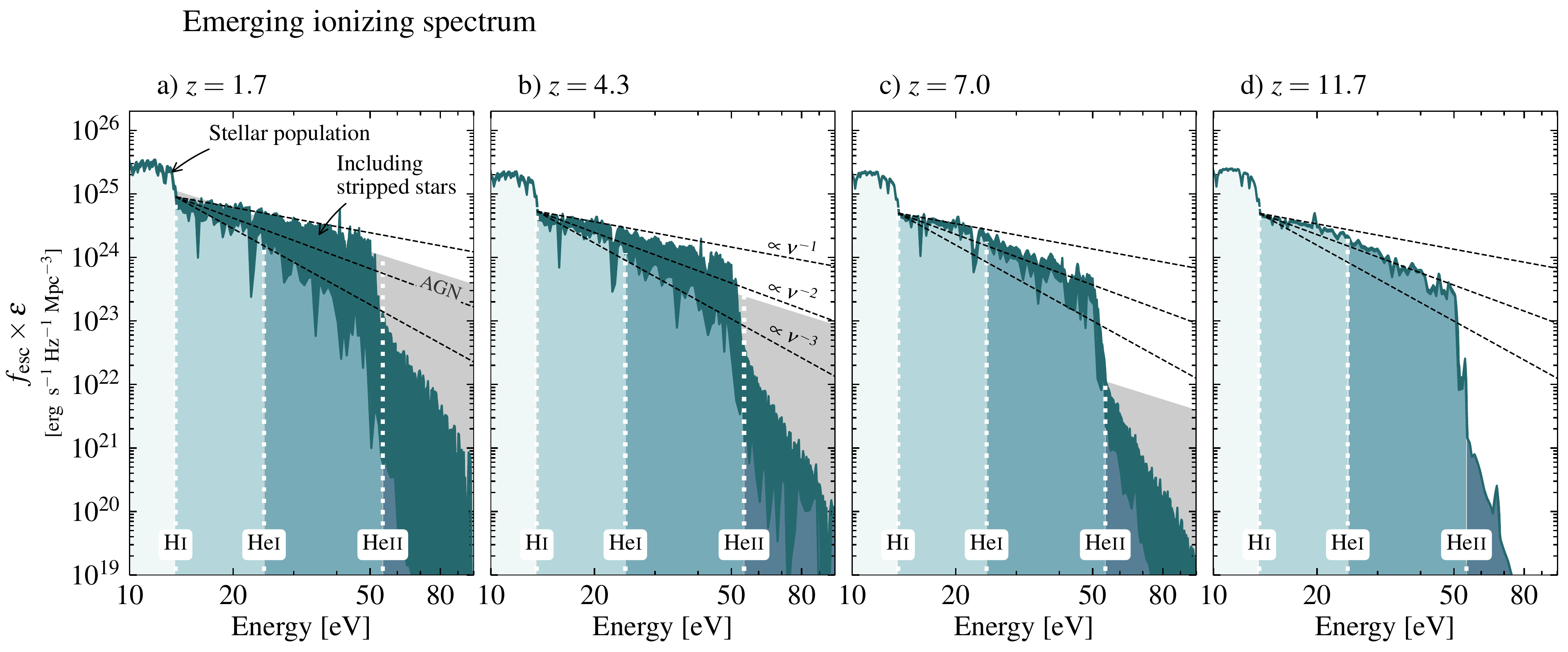}
\caption{Spectral energy distribution of the ionizing radiation that emerges into the intergalactic medium during cosmic evolution. We show the contribution from stellar populations in green shades and from AGN in gray. The contribution from stripped stars is highlighted in dark green. They cause the spectra to harden. The dashed lines show spectral slopes of $\alpha = -3$, $-2$, and $-1$, assuming $L_{\nu} \propto \nu ^{\alpha}$. The panels represent four snapshots during cosmic evolution, taken at $z = 1.7$, $4.3$, $7.0$, and $11.7$. Following the relation between average metallicity and redshift of \citet[][see also \secref{sec:stellar_populations}]{2017ApJ...840...39M}, the average metallicity at these times is $Z = 0.014$, 0.006, 0.002, and 0.0002 ($Z_{\odot} \equiv 0.014$, \citealt{2009ARA&A..47..481A}).
This figure adopts our standard model for the escape fractions, see \tabref{tab:models}. 
}
\label{fig:emissivity_spectrum}
\end{figure*}

%   - we show the total integrated spectrum of the sources that we considered over cosmic time in Figure
We show the integrated spectrum of the ionizing emissivity for four snapshots in the cosmic history in \figref{fig:emissivity_spectrum}, using our standard model. The figure displays the emerging emissivities of stripped stars, massive stars, and AGN at redshifts $z = 1.7$, 4.3, 7, and 11.7 and in units of erg~s$^{-1}$~Hz$^{-1}$ per co-moving Mpc$^3$. The emissivity, $\varepsilon$, is here defined as the ionizing spectrum integrated over all sources within a co-moving Mpc$^3$. We calculate the emissivity in a similar fashion as the production rate of ionizing photons (see \eqref{eq:ndotion_stars}) by multiplying the average wavelength-dependent luminosity of stellar populations, $L_{\nu}$, with the cosmic star-formation rate density. In \figref{fig:emissivity_spectrum}, we highlight the additional radiation that stripped stars contribute with using a dark green color. To create the figure, we used spectra for continuous star-formation during 1~Gyr. After such a long time, the spectrum has since long reached equilibrium in the ionizing wavelengths.

%   - FIGURE: SED at several points in redshift and showing the relative contributions + total (should show that stripped stars harden the spectra significantly)

% Main message of the figure
With their hard, ionizing spectra, stripped stars cause the total emerging emissivity from stellar populations to harden. 
% trends and numbers
The effect is primarily important for $z \lesssim 7$ and it becomes larger with decreasing redshift, as seen in \figref{fig:emissivity_spectrum}. We assume that the emerging emissivity roughly follows $\propto E^{\alpha}$ where $E$ is the photon energy and $\alpha$ is the spectral index. Then, at $z \sim 2$, stripped stars cause the emission to harden from $\alpha \sim -3$ to $\alpha \sim -1$ for photon energies smaller than about 50~eV. The effect is similar up to $z \sim 4$, and at $z = 7$, stripped stars cause the spectral index to increase by about $0.5$. %It is interesting to note that when stripped stars are included, the spectral hardness is similar at all redshifts, $\propto \nu^{-2}$. 
%The main contribution from stripped stars during the reionization of hydrogen was primarily extra \HI-ionizing photons. Their addition is included in \figref{fig:UVB}, however not easy to distinguish since we use logarithmic scale on the vertical axis.

% details
The panels of \figref{fig:emissivity_spectrum} show how the ionizing emission evolved as the Universe grew older. With lower redshift, the average metallicity increased, causing the emission from massive stars to soften from spectral indices of $\alpha = -2$ (panel d) to $\alpha =-3$ (panel a). However, when stripped stars are accounted for, this effect is washed out and the spectral index from stellar populations remains at $\alpha \geq -2$. 
% Comment on how this looks like for the optimistic and pessimistic models
The results for our high-\fesc and low-\fesc models are less pronounced but similar to those of the standard model. In the high-\fesc and pessimistic models, the stripped stars cause the spectral index to remain at $\alpha \sim -2$ over cosmic time. 

% subplots
The influence of AGN is small for the hardness of the \HI- and \HeI-ionizing spectrum. At these wavelengths, stellar populations dominate at all times in our models. However, the \HeII-ionizing emission is dominated by AGN for most of the cosmic evolution. Only at very high redshifts, $z \gtrsim 10$, stellar populations dominate the emission of \HeII-ionizing photons. It is likely that the contribution from stellar populations was larger than what our models predict at these high redshifts. The reason is that a spread of metallicities exist at each point in time, which allows for a larger contribution from stripped stars, while our models assume an average metallicity for each point in time. 
% Uncertainties (rotation?)
We also note that other sources of ionizing radiation may affect the hardness of the ionizing spectrum. Rotating massive stars could cause the \HI- and \HeI-ionizing emission to be slightly harder \citep[see e.g.,][]{2019ApJ...874..154D}, while accreting white dwarfs and X-ray binaries could affect the hardness of the \HeII-ionizing emission \citep{2015MNRAS.453.3024C, 2017ApJ...840...39M, 2019A&A...622L..10S}. %However, metallicity is an important ingredient for the hardness of the ionizing spectrum and metallicity variations over cosmic time or a spread of metallicity at each point in time could have a larger impact than what our simple model predicts. 

% Hardening of the ionizing spectrum by a wavelength dependent escape fraction
%The escape fraction is slightly wavelength dependent, as higher energy photons have longer mean-free paths and thus easier escapes the host galaxies \citep{2006agna.book.....O, 2017ApJ...845..111M}. The result is that the spectrum that emerges into the IGM is harder than the spectrum the stars produced. However, it is not yet clear how large the effect of a wavelength dependent escape fraction is. If photons escape through large holes in the galactic gas, the emerging radiation is not hardened since no gas is hindering their propagation. 
%We do not account for the wavelength dependence of the escape fraction, meaning that the ionizing emissivity is likely harder than what our models predict. 
Another effect that is expected to cause hardening of the emitted light is if the escape fraction is strongly dependent on the photon energy. The escape fraction is expected to be dependent on the photon energy since higher energy photons have longer mean-free paths \citep{2006agna.book.....O, 2017ApJ...845..111M}, but it is not yet clear how large the effect is. If photons escape through large holes in the galactic gas, the emerging radiation is not hardened since no gas is blocking their propagation.

%________________________________________________________________

% Describes how stripped stars affects the reionization of hydrogen and helium
\section{Reionization including stripped stars}\label{sec:reionization}

% This section, we show results from solving the equations in section 2 -- reason to have a figure in the first place
Here, we describe the results of our models for the cosmic evolution of the fraction of ionized hydrogen and helium. We focus in particular on the results from our standard model, but compare also to the results from the high-\fesc and low-\fesc models. First, we describe the effect of stripped stars on hydrogen reionization. Then, we discuss their impact on helium reionization.

\subsection{Hydrogen reionization}

% FIGURE: Volume filling factor of ionized element over cosmic time
\begin{figure*}
\centering
\includegraphics[width=.65\textwidth]{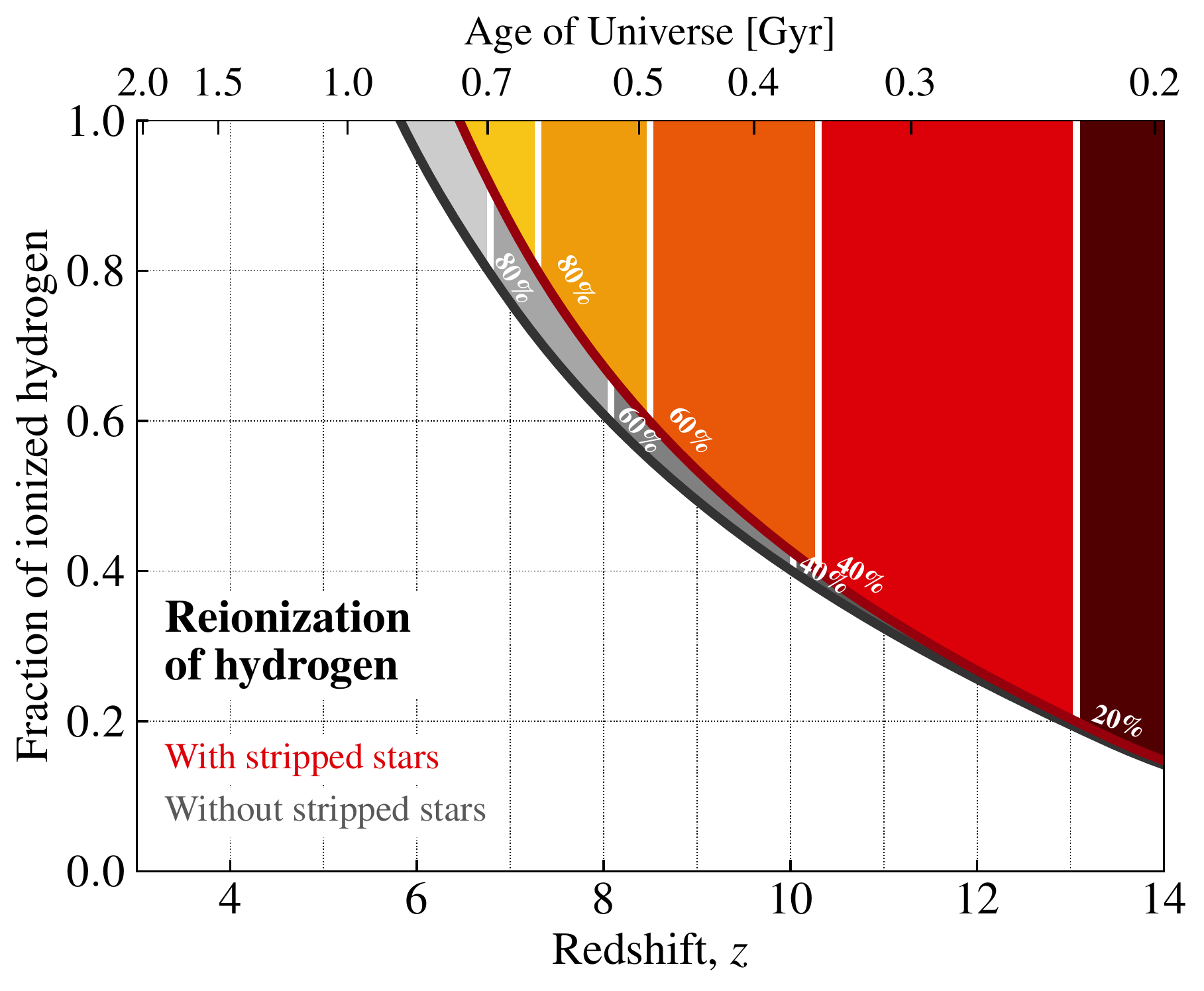}
\caption{The reionization of hydrogen shown as the cosmic evolution of the volume filling factor of ionized hydrogen in the IGM, $x_{\HII}$. We show the evolution of $x_{\HII}$ in our standard model (see \tabref{tab:models}). We use red shades for the version in which stripped stars are included and gray shades for when they are not included. We mark the times when 20\%, 40\%, 60\%, and 80\% of the hydrogen in the IGM is ionized using white lines and labels. Stripped stars cause reionization to occur about 100~Myr earlier than when only single stars and AGN are considered.}
\label{fig:vol_fill}
\end{figure*}

% Table with the results for when reionization is occurring
\begin{table}
\centering
\caption{The redshifts at which hydrogen reionization is completed for our four models that differ in their assumption for the escape fraction of ionizing photons (see \tabref{tab:models}). In columns two and three we show the redshifts when reionization is completed for when stripped stars are not included ($z_{\text{\, EoR, no strip}}$) and for when stripped stars are included ($z_{\text{\, EoR, strip}}$), respectively. In the fourth column we show the fraction of the total amount of ionizing photons that stripped stars contributed to the reionization of hydrogen, $f_{\text{strip}}$.}
\label{tab:fesc}
\begin{tabular}{lccc}
\toprule\midrule
Model & $z_{\text{\, EoR, no strip}}$ & $z_{\text{\, EoR, strip}}$ & $f_{\text{strip}}$ \\
\midrule
%\textbf{Standard}      & 5.8258258258258255  &  6.456456456456457  \\
%\textbf{Optimistic}     &  6.246246246246246 & 6.546546546546547 \\
%\textbf{Pessimistic}   &  4.414414414414415 & 4.714714714714715  \\ 
%\textbf{Extreme}     & 4.414414414414415 & 5.105105105105105 \\
Standard      &  5.83 & 6.46 &  21.2\% \\
High-\fesc     &  6.25 & 6.55 & 16.1\% \\
Low-\fesc   & 4.41 & 4.71 & 21.1\% \\ 
Combination   & 4.41 & 5.11 & 42.3\% \\
\bottomrule
\end{tabular}
\end{table}

% Present figure - what is inside? 
We show how the volume filling factor of ionized hydrogen, $x_{\HII}$, increases with decreasing redshift in \figref{fig:vol_fill}. 
% describe figure
% Main message from the differential equation solutions -- stripped stars could help H-reionization occur on time
We find that hydrogen reionization occurs about 100~Myr earlier when stripped stars are included, which corresponds to about 10\% of the total age of the Universe at the time. In our standard model, \HI-reionization occurs at $z \sim 5.8$ if stripped stars are not included, but at $z \sim 6.5$ if they are. 
% Trends - similar at first, very different at later stages
%   - quantify: 75% reionized when the other model is 100%
The impact of stripped stars is small at early times, but they become important as hydrogen reionization progresses. At redshift $z = 12$, about 25\% of the Universe is reionized independent on whether stripped stars are included, but when hydrogen is completely reionized in the model including stripped stars, only 85\% of the IGM is reionized if stripped stars are not included. 
%In both models, the reionization starts slowly and it takes about 500~Myr for half of the Universe to be reionized. However, the completion is faster and the Universe goes from half to completely reionized in about 200~Myr. 
In \tabref{tab:fesc}, we display the redshifts at which hydrogen reionization is completed for our four models and for when stripped stars are accounted for and when they are not. The table shows that the effect of the escape fraction is large for the time of reionization. However, in all models, stripped stars give rise to earlier reionization compared to if they are not included.

% % % % % % % Relative importance of the different sources

% Relative importance
To evaluate the total relative importance of the different ionizing sources to the reionization of hydrogen, we integrate the individual contributions to the cosmic emission rate of \HI-ionizing photons from the early Universe up to the end of reionization. The resulting contribution from stripped stars to the budget of \HI-ionizing photons prior and up to the reionization is about 20\% in our standard model. The effect of the choice of the escape fraction is visible when comparing the relative contribution from stripped stars between our models. In the combination model, the stripped stars have a much larger escape fraction than massive stars, which results in that their contribution is as high as about 40\%. On the other hand, in our high-\fesc model, the escape fraction for stripped stars is the same as in our combination model, but the escape fraction for massive stars is lower. The result is that the reionization occurs earlier and the contribution from stripped stars is smaller ($\sim 15$\%, see \tabref{tab:fesc}).

% Figure showing the relative contributions to the H-ionizing photon budget. 
\begin{figure}
\centering
\includegraphics[width=\hsize]{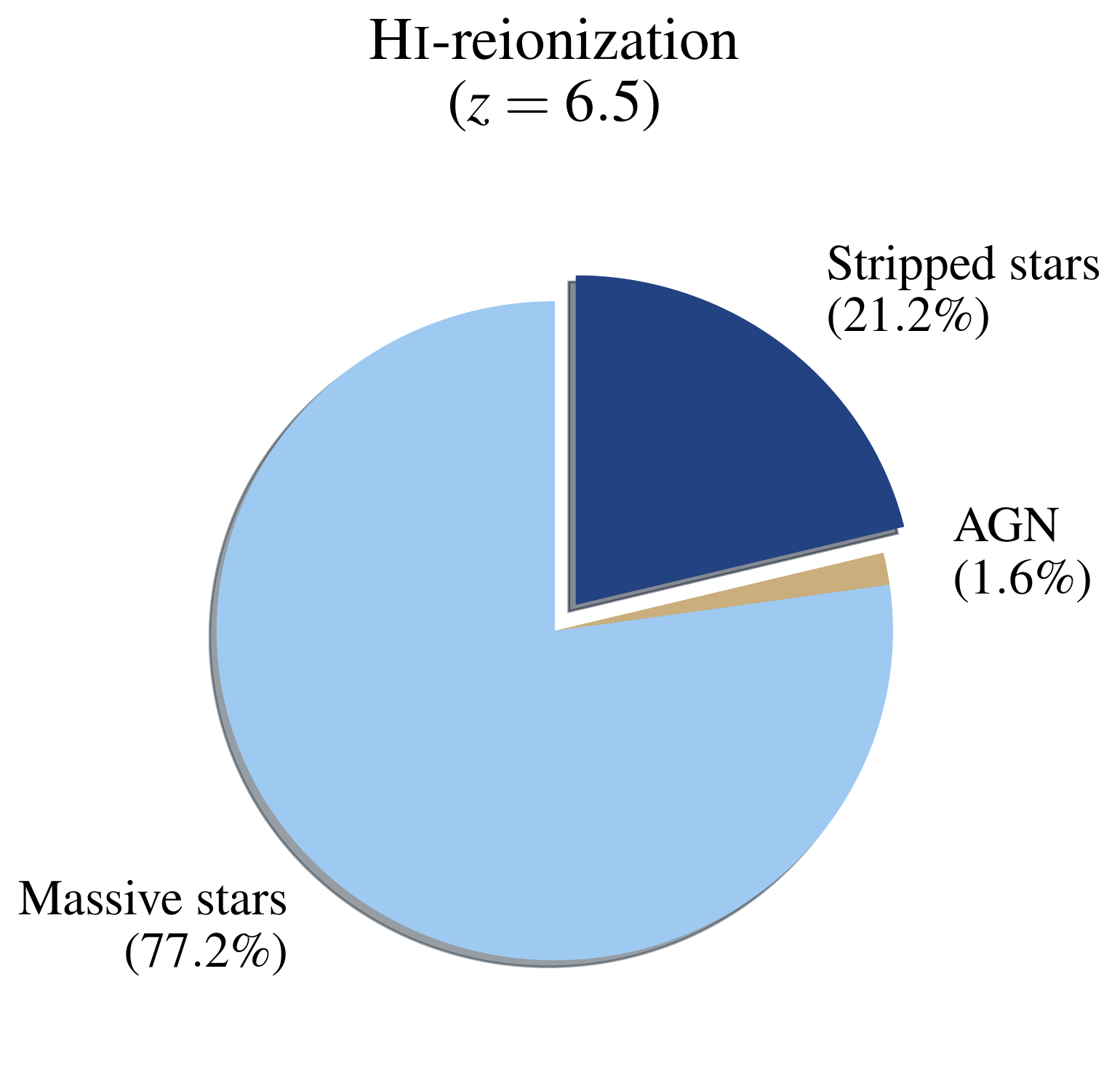}
\caption{The relative contribution of ionizing photons from stripped stars (dark blue), massive stars (light blue), and AGN (beige) to the photon budget that caused hydrogen reionization. Stripped stars contributed with about 20\% of the \HI-ionizing photons, see \secref{sec:reionization}. This figure shows the results from our standard model, for which reionization is complete at $z = 6.5$ if stripped stars are included (see \tabreftwo{tab:models}{tab:fesc}). }
\label{fig:pie_summary}
\end{figure}

% Discuss the relative importance
\figref{fig:pie_summary} shows the relative contribution from the different ionizing sources to the budget of \HI-ionizing photons, measured from the early Universe up until the completion of reionization at $z = 6.5$ for our standard model. The figure shows that stripped stars played an important role as contributors with ionizing photons during the reionization of hydrogen, providing about 20\% of the ionizing photons. Massive stars were the most important, producing about 75\% of the ionizing photons. In our models, AGN are less important as sources of \HI-ionizing photons and produce only a few percent of the photons that reionized the Universe. 
We also show the evolution of the relative contributions from the different ionizing sources to the total number of produced \HI-ionizing photons in \figref{fig:rel_Nion}a. This figure shows that the importance of stripped stars increased with the age of the Universe. %However, this effect is dependent on the choice of the escape fraction.

% Is it possible that quasars played a more important role? 
The contribution from AGN has been debated and several models consider a higher number density of AGN at earlier times, which leads to a larger contribution to the hydrogen reionization \citep[e.g.,][]{2015A&A...578A..83G, 2015ApJ...813L...8M}. However, higher emission from AGN at early times leads to earlier reionization of helium than what has been observed because of the hard spectra of AGN. We, therefore, do not consider any model where the contribution from AGN is larger.

% Comparison with BPASS
%   - BPASS emission rate of ionizing photons is X times higher -- see appendix
%   - the reason it is higher is because of chemically homogeneous stars, optimistic approach
%   - result is that BPASS models can produce reionization with lower escape fractions, but numerical models suggest it even happens earlier than expected. 
In the code \code{BPASS} \edit{\citep[][version 2.2.1, \textit{Tuatara}]{2017PASA...34...58E, 2018MNRAS.479...75S}}, the inclusion of binary products results in \edit{a few percent up to 30 percent} higher emission rates of hydrogen ionizing photons depending on the considered metallicity (see the yields of ionizing photons presented in \figref{fig:Iion}). This is consistent with what we predict for stripped stars at solar metallicity, which produce about 5\% of the total number of emitted ionizing photons from stellar populations \edit{\citep[see also][]{2016MNRAS.456..485S}}. At lower metallicity, \code{BPASS} predicts higher yields of \HI-ionizing photons than what we predict stripped stars can be responsible for. The individual yields of ionizing photons for different types of stars are not available in the \code{BPASS} models. However, the most likely reason for their higher yields is that for $Z \leq 0.004$, \code{BPASS} accounts for other binary products than stripped stars, and especially for efficient rotational mixing and subsequent chemically homogeneous evolution of low-metallicity stars that accreted material or merged with its companion star \citep[][\edit{see also \citealt{2007A&A...465L..29C}}]{2017PASA...34...58E}. In consequence, \code{BPASS} predicts complete reionization for low escape fractions within the observed timescale or even gives rise to early reionization \citep[$\fesc \sim 5-10$\%,][see also \citealt{2016MNRAS.459.3614M} \edit{ and \citealt{2016MNRAS.458L...6W}, but note that these authors all used an earlier version of \code{BPASS}, which had a higher binary fraction than in the version \textit{Tuatara}}]{2018MNRAS.479..994R}. This shows that the effect of chemically homogeneous stars is significant, \edit{with important implications for cosmic reionization and the spectra of stellar populations},if indeed rotational mixing is as efficient as these models predict. \edit{Although several studies have claimed circumstantial evidence \citep[e.g.,][]{2009A&A...495..257M, 2015A&A...581A..21H, 2018A&A...611A..75S, 2019ApJ...880..115A}, direct, solid proof for chemically homogeneous evolution is still missing.}
%\edit{Unfortunately, direct, solid evidence for chemically homogeneous evolution is still missing, although several claims for circumstantial evidence have been made \citep[e.g.\ ][]{2009A&A...495..257M, 2015A&A...581A..21H, 2018A&A...611A..75S, 2019ApJ...880..115A}.}

% % % % %    Helium reionization    % % % % % 
%
\subsection{Helium reionization}

% Brief: stripped stars did not play a very important role for the reionization of helium (numbers, no figure)
%   - time of helium reionization not affected
%   - relative contribution with helium ionizing photons is small (~2 %)
All our models indicate that stripped stars contributed in total with $\lesssim 3$\% of the photons that fully reionized helium. This means that the time of helium reionization is not affected by stripped stars. Our models reach complete helium reionization at $z \sim 3$ and the main producer of the \HeII-ionizing photons that caused helium reionization in our models is AGN.

% FIGURE showing the relative contribution as evolution with redshift
%   - highlight early redshifts, maybe entire evolution?
\begin{figure}
\centering
\includegraphics[width=0.85\hsize]{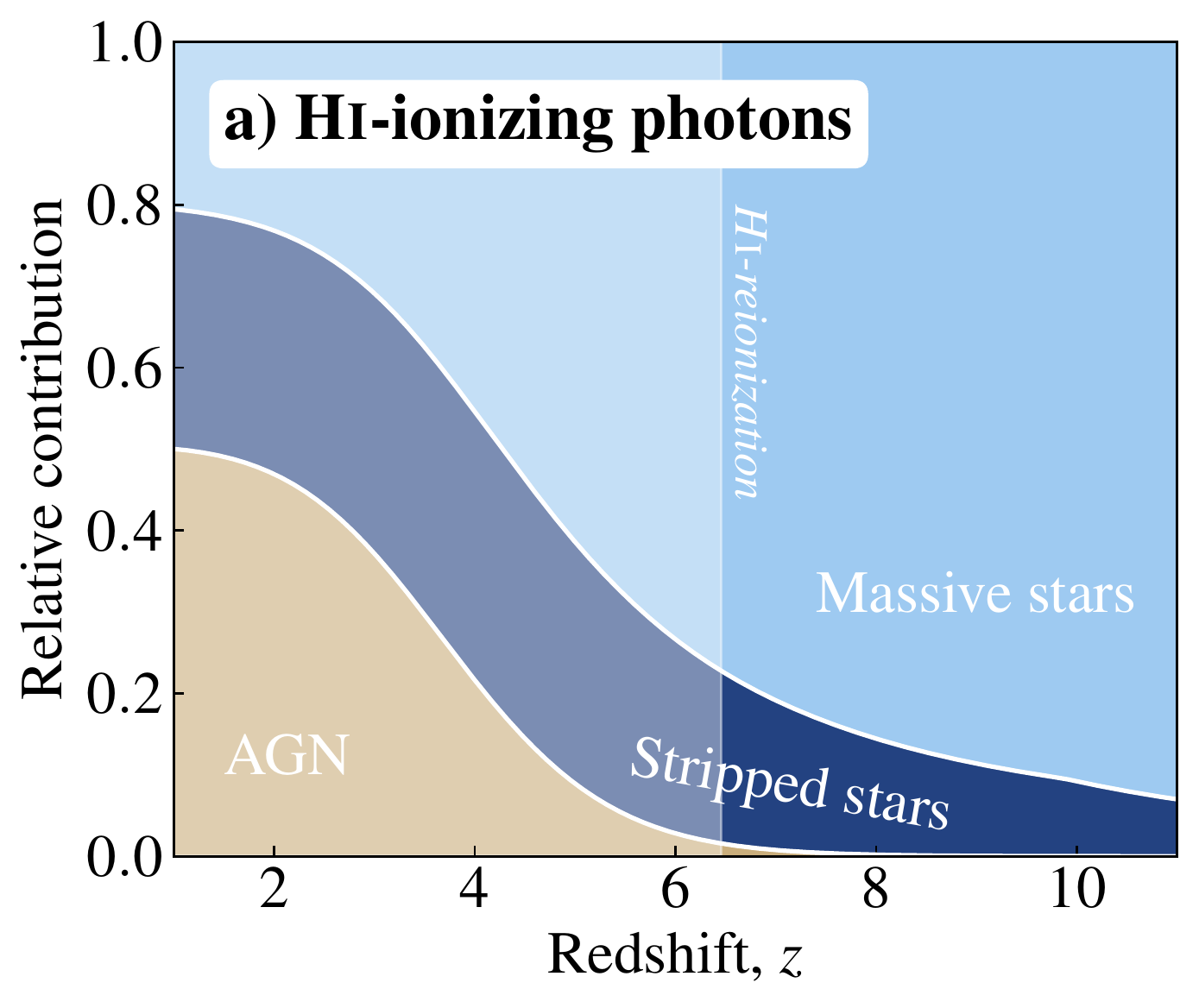} \\
\includegraphics[width=0.85\hsize]{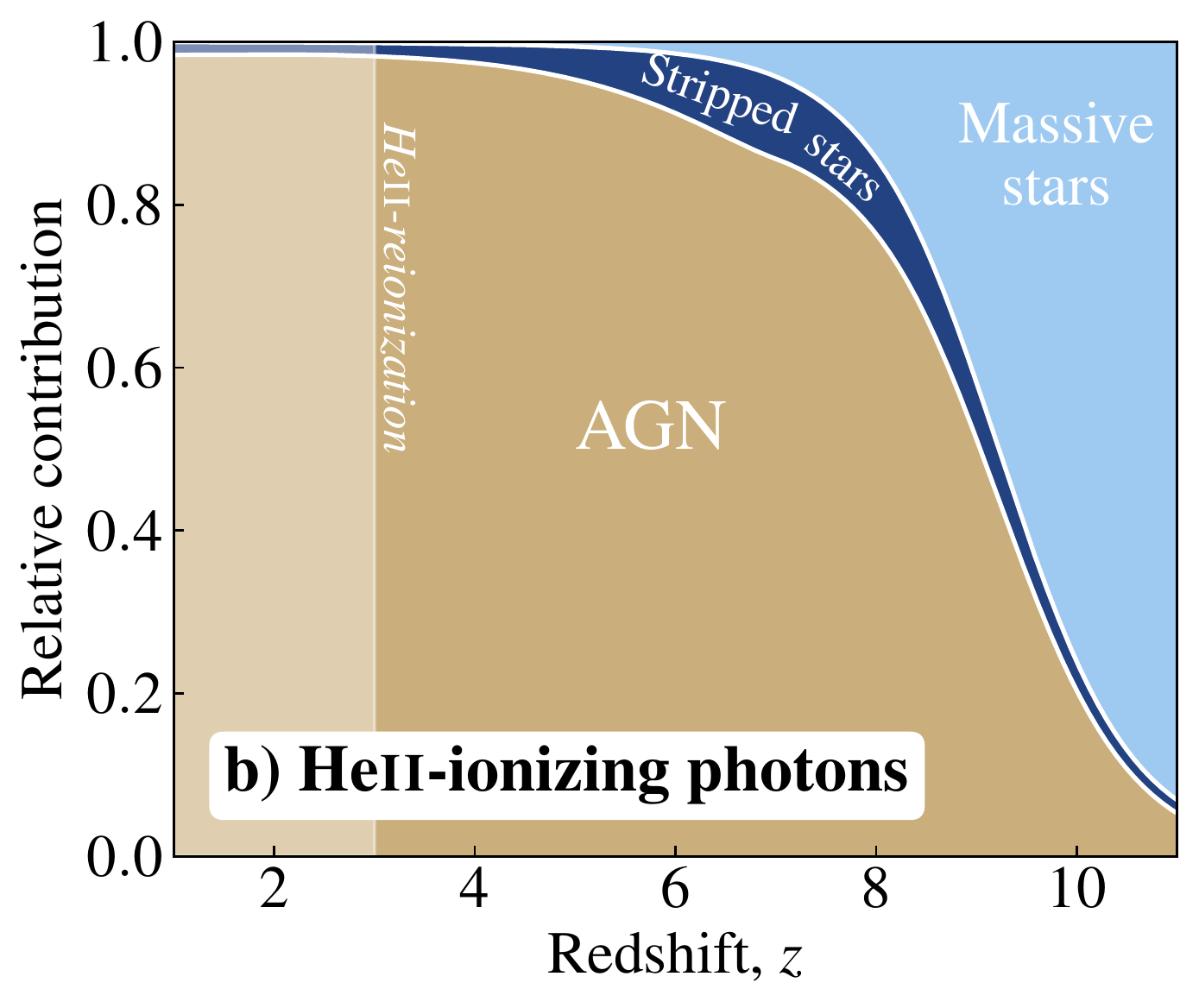}
\caption{The cosmic evolution of the relative contribution from stripped stars (dark blue), massive stars (light blue), and AGN (beige) to the budget of \HI-ionizing photons (panel a) and \HeII-ionizing photons (panel b). We label the time of hydrogen and helium  reionization and shade the times our model predicts that the Universe is reionized. This figure shows the results from our standard model; the trend in massive stars is in part due to the strongly redshift-dependent \fesc (see \tabref{tab:models}). }
\label{fig:rel_Nion}
\end{figure}

% Where they might have mattered was at early times, when they actually could have been an important source of HeII ionizing photons
%   - describe figure: main message
%   - what could be the impact of these early HeII-ionizing photons?
We show the relative contributions to the budget of \HeII-ionizing photons from the different ionizing sources that we consider in \figref{fig:rel_Nion}b. The figure shows that the contribution from stellar populations was dominating before $z \sim 10$ and between $z \sim 6-8$, stripped stars were responsible for about 10\% of the \HeII-ionizing photons that reached the IGM. This means that the early emission of \HeII-ionizing photons could have resulted in heating of the IGM before the reionization of helium was complete. In our standard model, the reionization of helium starts around $z = 5$, when slightly more than 10\% of the intergalactic helium has been fully ionized.

%________________________________________________________________
% Effect from stripped stars on observables during EoR
% subsections include: heating of IGM, cooling of IGM, elements in IGM
%\section{Discussion}\label{sec:discussion}
\section{Observable consequences}\label{sec:discussion}

% Harder ionizing spectra needed to reproduce the strong absorption of CIV that is observed
\subsection{Highly ionized absorbers in the IGM}

% Background: highly ionized species have been observed in the IGM at z~6. The metals likely come from galactic outflows, which explains that the metals are detected in small and dense clumps of the IGM. The metals are presumable ionized by the emission from the galaxy, and we can see the lines in absorption from studying the spectra of background quasars. 
In \secref{sec:SED_results}, we showed that stripped stars harden the ionizing radiation that emerges from galaxies (\figref{fig:emissivity_spectrum}), which could produce significant effects on the observable ionization structure of the IGM. This is particularly interesting because high-ionization species such as \CIV and \SiIV have been observed in the IGM at high redshift \edit{\citep[e.g.,][]{2006MNRAS.371L..78R, 2009MNRAS.395.1476R, 2013MNRAS.435.1198D, 2018MNRAS.475.4717D}. }
The metals in the IGM are likely expelled from galaxies via large-scale outflows, which explains why they are detected in small and dense clumps that could indicate the locations of galaxies. The intergalactic metals are then ionized by radiation emitted by the galaxies, reaching various levels of ionization depending on the hardness and intensity of the emerging ionizing radiation.

% There is more CIV absorbers observed compared to what models predict. A reason for this could be either harder ionizing spectra or more luminous ionizing spectra compared to what the models assume. 
The observed abundance of \CIV at high redshifts is in some cases unexpectedly high compared to what cosmological simulations predict \citep[e.g.,][]{2016MNRAS.459.2299F, 2018MNRAS.475.4717D}. A reason could be that the emission from galaxies is harder than what is assumed for the galaxy spectra for the \HI- and \HeI-ionizing emission at energies between $1-4$~Rydberg \citep{2013MNRAS.435.1198D}.

% Stripped stars harden the spectra and could thus help increase the amount of CIV surrounding galaxies -- understand how outflows work, do they happen on timescales correlated with star-formation? 
%Stripped stars harden the ionizing radiation that emerges from galaxies and their effect on the intergalactic metals is therefore an interesting topic. 
Stripped stars emit sufficiently hard ionizing radiation to ionize both \SiIII to \SiIV and \CIII to \CIV \citepalias[see Fig.~7 of][]{2019A&A...629A.134G}.
%The ionization thresholds of \SiIII to \SiIV and \CIII to \CIV are both within the \HeI-ionizing energy range at 33~eV and 48~eV, respectively \citepalias[see e.g., Fig.~7 of][]{2019A&A...629A.134G}. Stripped stars contribute significantly with radiation within this energy range, as seen in \figref{fig:emissivity_spectrum}. 
\figref{fig:emissivity_spectrum} shows that our models predict that stripped stars boost the flux at these photon energies by a factor of a few up to ten, depending on the metallicity of the stellar population. % Which test is needed? Can we do a quick back-of-envelope estimate of how much they may help? Is there a relation between Omega_{CIV} and spectral hardness? 
However, detailed radiative transfer modeling is needed to accurately determine the influence from stripped stars on the ionization state of the IGM. \edit{We also note that other sources of hard, ionizing radiation, such as accreting compact objects \citep[e.g.,][]{2003ApJ...592..884D, 2013ApJ...764...41F, 2015MNRAS.453.3024C} or post-AGB stars \citep[e.g.,][]{2019AJ....158....2B}, may play a role in the creation of highly ionized elements in the IGM.}

\edit{
We note that the absorption lines discussed here are different from the strong emission lines observed in some star forming galaxies, for example \OIII and \HeII \citep{2015MNRAS.454.1393S, 2017MNRAS.472.2608S, 2019ApJ...878L...3B}. To make predictions for these one would need to carefully model the nebular spectrum, which is beyond our current scope.  
}

% Is there any relation that could give how likely it is that certain elements are in certain ionization stages? 
%\citet{2016MNRAS.459.2299F} modeled the intergalactic metal absorption features using the ionizing emission as modeled by \citet{2012ApJ...746..125H} (assuming the single star models by \citet{2003MNRAS.344.1000B} for stellar populations). 

% Impact on the temperature in the IGM
\subsection{Temperature of the intergalactic medium}

% Short subsection
%   (1) describe why stripped stars could impact the temperature of the IGM
%   (2) discuss how that difference potentially can be detected. (phrase it as stripped stars change the temperature, not stripped stars can be inferred via the temperature.)

% The IGM is heated depending on the hardness of the ionizing spectrum
%   - harder spectrum -> hotter IGM
%   - IGM cools relatively quickly
During reionization, the IGM was heated by excess photon energy after each ionization of individual atoms. The gas simultaneously cooled through line excitations and collisional ionization from to photoelectrons \citep{1994MNRAS.266..343M}. Harder spectra of the ionizing sources leave, therefore, imprints on the IGM in terms of higher temperatures. After reionization is complete, the IGM cooled again, reaching temperatures of about $7,000 - 10,000$~K at $z\sim 5$ \citep{2011MNRAS.410.1096B, 2014MNRAS.441.1916B, 2017PhRvD..96b3522I}.

% Stripped stars make the spectrum harder -> can impact the temperature of IGM
%   - d'Aloisio+18 gives a good estimate for how much it might get affected: 1-5 kK depending on speed of ionization fronts and hardening of the spectrum
With the hardening of the ionizing spectrum when stripped stars are considered, the IGM may reach higher temperatures depending on the speed with which the ionization fronts move \citep[faster fronts lead to larger temperature increases, ][]{2019ApJ...874..154D}. Following Fig.~2 of \citet{2019ApJ...874..154D} and assuming that the ionization fronts move with speeds between $10^3$ and $10^4$\kms, we estimate that stripped stars could cause an increase of the IGM temperature of $\sim$ 1,000 up to $\sim$5,000~K since they harden the ionizing emission and cause significant changes to the spectral index (see \figref{fig:emissivity_spectrum}). 

% The higher temperature of the IGM may mean that 
%   - Lya forest is slightly broader (refs)
%   - Lya emission from ionization fronts is stronger (futuristic, SPHEREx)
Stripped stars may therefore affect observables for the temperature of the IGM. The extra heating from the hard spectra of stripped stars would cause lines in the Ly$\alpha$ forest to be more thermally broadened.  Temperature measurmeents from the forest have been performed to redshifts as high as $z\approx 6$ \citep{2007MNRAS.374..493B, 2012MNRAS.419.2880B,2013PhRvD..88d3502V, 2019ApJ...872..101B}. A precise understanding of the timing of reionization is likely required to be able to use a measurement of temperature to probe the hardness of the sources' spectra as the post-reionization temperature that is measured in the forest depends also on when gas was ionized (affecting how much it is able to cool; \citealt{2016MNRAS.460.1885U}). Such an understanding may be possible in combination with future cosmic microwave background and redshifted 21cm observations \citep{2016ARA&A..54..313M}. 
%\YG{Comment on the recent better measurement}

The ionization fronts during reionization cool via collisionally excited Ly$\alpha$ emission. Thus, with a higher IGM-temperature, the line emissivity of Ly$\alpha$ is expected to increase \citep{2008ApJ...672...48C, 2018ApJ...863L...6V}. We estimate that the increased IGM temperature that stripped stars are responsible for give rise to an increased Ly$\alpha$ emissivity of about a factor of a few up to ten \citep[cf.\ Fig.~1 of][see also \citealt{2016MNRAS.457.3006D}]{2008ApJ...672...48C}. 
However, the Ly$\alpha$ emission from cooling ionization fronts during the reionization is hard to detect \citep{2013ApJ...763..132S, 2014ApJ...786..111P, 2016MNRAS.457.3006D}. This signal then needs to be distinguished from galactic Ly$\alpha$ emission \citep[however, see][]{2016MNRAS.463.3078C}, but there may be unique signatures that allow it to be extracted \citep{2018ApJ...863L...6V}. The weak signal from the reionizing IGM is predicted to be detectable with instruments equipped with higher sensitivity than current, such as the future space mission SPHEREx \citep[][launch expected in 2023]{2017arXiv170909066K}.

%________________________________________________________________
% Summary and Conclusions:
%  importance of stripped stars for each EoR
%  important uncertainties
%  possible effect on observables
%  outlook
\section{Summary \& Conclusions}\label{sec:summary_conclusions}

% Summary
%   - Simple analytical model: just to test the importance of the ionizing radiation from stripped stars on the reionization. Based on detailed spectral and evolutionary modeling of stars. Our model is approximative, but it proves a point. Better model would be to include the yields we present in cosmological simulation boxes.
\edit{With their high temperatures and late formation times, stars stripped of their hydrogen-rich envelopes during interaction in binaries have been predicted to play an important role during the reionization. 
We have estimated their contribution to the budget of photons that caused cosmic reionization of hydrogen and helium. We used detailed spectral models custom-made for these stripped stars \citepalias{2018A&A...615A..78G}. }
We combined the radiative contribution from stripped stars to realistic stellar and AGN populations \citepalias{2019A&A...629A.134G} with a \edit{commonly used,} simple method to calculate the cosmic evolution of the volume filling factors of ionized hydrogen and helium. 
%   - We include escape fraction, recombination timescale and use together with cosmic SFH on top of Starburst99 and AGN emissivity. We also assume clumping of the IGM and the primordial density 
We accounted for the cosmic star-formation history, the recombination timescales, the intergalactic density of hydrogen and helium, and assumed clumping of the intergalactic medium. %Our model accounts for the effects of stars stripped in binaries, but neglects the impact of other binary products, which could also affect the ionizing emission from stellar populations. 
We have considered four models, which differ from each other by the different assumptions of the escape fraction. With these various approaches for the uncertain escape fraction, we hope to have bracketed the actual conditions between stellar populations and intergalactic space.

% Conclusions
\edit{
In agreement with previous studies, we have found that stars stripped in binaries likely had important effects on cosmic reionization. However, we have taken a different approach than earlier work on the effect of interacting binaries on cosmic history \citep[cf.][]{2016MNRAS.459.3614M, 2018MNRAS.479..994R}. 
% focus on stripped stars - understand only their role
% detailed models - understand how certain our predictions are
By carefully implementing detailed stellar evolutionary and atmosphere models especially made for stripped stars in a model for reionization, we have made it possible to: (1) follow a single type of binary product and understand their impact on the evolving Universe, and (2) better understand which predictions for them are robust and which are uncertain.
% simple model, but instructive, more sophisticated models could do it somewhat better, maybe especially fesc
Our reionization model is simple. More sophisticated cosmological simulations would aid in understanding details about the role of stripped stars during reionization. In particular, such simulations could self-consistently model the escape of ionizing photons, which is a critical uncertainty in our analysis. 
% with our simple model, we could identify potential observable traces stripped stars left over cosmic history.
However, with the simple model we employ, the effect originating from individual ionizing sources is easy to distinguish. This has allowed us to identify possibly observable traces that stripped stars left in cosmic history.
}

\edit{Below, we summarize the conclusions we have found.}
%We summarize our conclusions as follows:
\begin{enumerate}

\item Stripped stars cause the emerging spectra from galaxies to harden. At solar metallicity, the slope of the ionizing part of the spectrum can harden from a spectral index of $-3$ to $-1$ if stripped stars are included. The effect decreases with lower metallicity. \\

\item Our standard model suggests that stripped stars provided $\sim 20\%$ of the photons that reionized hydrogen in the Universe. This estimate is sensitive to our assumptions about the escape fraction, however, the uncertainties in binary interaction and in the emission rates of \HI-ionizing photons from stripped stars are expected to be small. \\

\item The \HeII-ionizing emission from stripped stars is not sufficient to compete with AGN and significantly impact the complete reionization of helium, even given the large uncertainties in our models about the absorption of \HeII-ionizing photons in a wind. However, if most of these photons are not absorbed, the stripped stars could have contributed significantly to the initial part of \HeII\ reionization at $z \sim 6-8$. \\

\item The harder ionizing radiation that stripped stars introduce to the spectra of galaxies could give rise to high ionization levels of high-redshift circumgalactic gas. This could potentially help explain the unexpectedly high abundance of \CIV that has been detected in dense regions at $z \sim 6$ \citep[see e.g.,][]{2013MNRAS.435.1198D}. \\

\item The harder spectra that we expect from stellar populations because of the presence of stripped stars likely lead to an increased temperature of the intergalactic medium during the reionization of hydrogen. We estimate that accounting for stripped stars leads to an increase of 1,000 to 5,000~K in the temperature of the IGM at the time of ionization. Current high-redshift temperature measurements using the Ly$\alpha$ forest likely could not detect such differences. The increased temperature could also lead increased emissivity of Ly$\alpha$ from cooling ionization fronts, which may be detectable with futuristic intensity mapping surveys. 
\end{enumerate}

% Outlook: we modeled stripped stars, what next? 
\noindent In this study, we have focused on the ionizing contribution from stripped stars during the reionization of hydrogen and helium. 
Performing detailed analysis of other products of binary interaction is important to improve our understanding further. However, the models we have presented for stripped stars already provide new clues on the nature of ionizing photons on cosmological scales. 
In combination with careful nebular modeling, our predictions may provide observable signatures in the large amounts of spectra anticipated from James Webb Space Telescope \citep[JWST,][]{2006SSRv..123..485G}, which would allow probing the role of interacting binaries across cosmic time.

% ______________________________________________________________________________

\begin{acknowledgements}
% Amy Secunda, Renyue Cen, Pratika Dayal, Anne Hutter, Dan Stark, Tony Piro, Jim Fuller, Danielle Berg, Gwen Rudie, Allison Strom, Claus Leitherer, JJ Eldridge, Elizabeth Stanway
\edit{The authors thank the anonymous referee for useful comments that helped improve the manuscript. 
YG acknowledges the funding from the Alvin E.\ Nashman fellowship for Theoretical Astrophysics. }
SdM acknowledges funding by the European Union's Horizon 2020 research and innovation programme from the European Research Council (ERC), Grant agreement No.\ 715063), and by the Netherlands Organisation for Scientific Research (NWO) as part of the Vidi research program BinWaves with project number 639.042.728. 
\edit{EZ acknowledges support from the Federal Commission for Scholarships for Foreign Students for the Swiss Government Excellence Scholarship (ESKAS No.\ 2019.0091) for the academic year 2019-2020.
This work made use of v2.2.1 of the Binary Population and Spectral Synthesis (\code{BPASS}) models for comparison purposes as last described in \citet{2017PASA...34...58E} and \citet{2018MNRAS.479...75S}.}
\end{acknowledgements}

%________________________________________________________________

% References, style A&A
\bibliographystyle{aa.bst}
\bibliography{references_bin.bib}

\appendix

% This appendix is meant to discuss how much it matters if a different model is used
\section{Yields of ionizing photons for a range of metallicity}\label{app:model_variations}

% Make a table with I_ion values computed for stripped stars
\begin{table*}
\centering
\caption{The yields of \HI-, \HeI-, and \HeII-ionizing photons (\Iionz, \Iiono, and \Iiont respectively). We show the contribution from stripped stars and from massive stars for a range of metallicities. }
\label{tab:Iion}
{\small
\begin{tabular}{lllll}
\toprule\midrule
Metallicity & Stripped stars & Starburst99 & BPASS (single) & BPASS (binary) \\ 
\midrule
\multicolumn{5}{l}{\textbf{Yields of \HI-ionizing photons}, ${I}_{\mathrm{ion,\HI}}$ in $10^{52}$ s$^{-1}$ (M$_{\odot}$\, yr$^{-1}$)$^{-1}$}\\ 
0.014 & 0.78 & 14.06 & 14.19 & 14.66\\ 
0.006 & 0.86 & 14.37$^{\dagger}$ & 16.43 & 18.99\\ 
0.002 & 1.05 & 20.92 & 20.12 & 24.36\\ 
0.0002 & 0.91 & 21.15$^{\ddagger}$ & 24.8$^{*}$ & 31.53$^{*}$\\ 
\midrule 
\multicolumn{5}{l}{\textbf{Yields of \HeI-ionizing photons}, ${I}_{\mathrm{ion,\HeI}}$ in $10^{52}$ s$^{-1}$ (M$_{\odot}$\, yr$^{-1}$)$^{-1}$}\\ 
0.014 & 0.41 & 2.08 & 2.55 & 2.76\\ 
0.006 & 0.43 & 2.54$^{\dagger}$ & 3.4 & 4.2\\ 
0.002 & 0.51 & 4.37 & 4.74 & 6.35\\ 
0.0002 & 0.26 & 5.02$^{\ddagger}$ & 7.66$^{*}$ & 10.62$^{*}$\\ 
\midrule 
\multicolumn{5}{l}{\textbf{Yields of \HeII-ionizing photons}, $\log_{10} ({I}_{\mathrm{ion, \HeII}}$/(s$^{-1} ($M$_{\odot}$\, yr$^{-1}$)$^{-1}$)}\\ 
0.014 & 49.1 & 48.6 & 49.6 & 49.5\\ 
0.006 & 49.1 & 49.5$^{\dagger}$ & 49.7 & 49.6\\ 
0.002 & 49.1 & 49.1 & 50.0 & 50.2\\ 
0.0002 & 46.9 & 48.9$^{\ddagger}$ & 49.5$^{*}$ & 50.4$^{*}$\\ 
\bottomrule
\end{tabular}

}
\tablefoot{$^{\dagger}$ quantity calculated for models with $Z = 0.008$, $^{\ddagger}$ quantity calculated for models with $Z = 0.001$, $^{*}$ quantity calculated for models with $Z = 0.0001$. ($Z_{\odot} \equiv 0.014$, \citealt{2009ARA&A..47..481A}.)}
\end{table*}

%  Figure with Iion
\begin{figure*}
\centering
\includegraphics[width=0.33\textwidth]{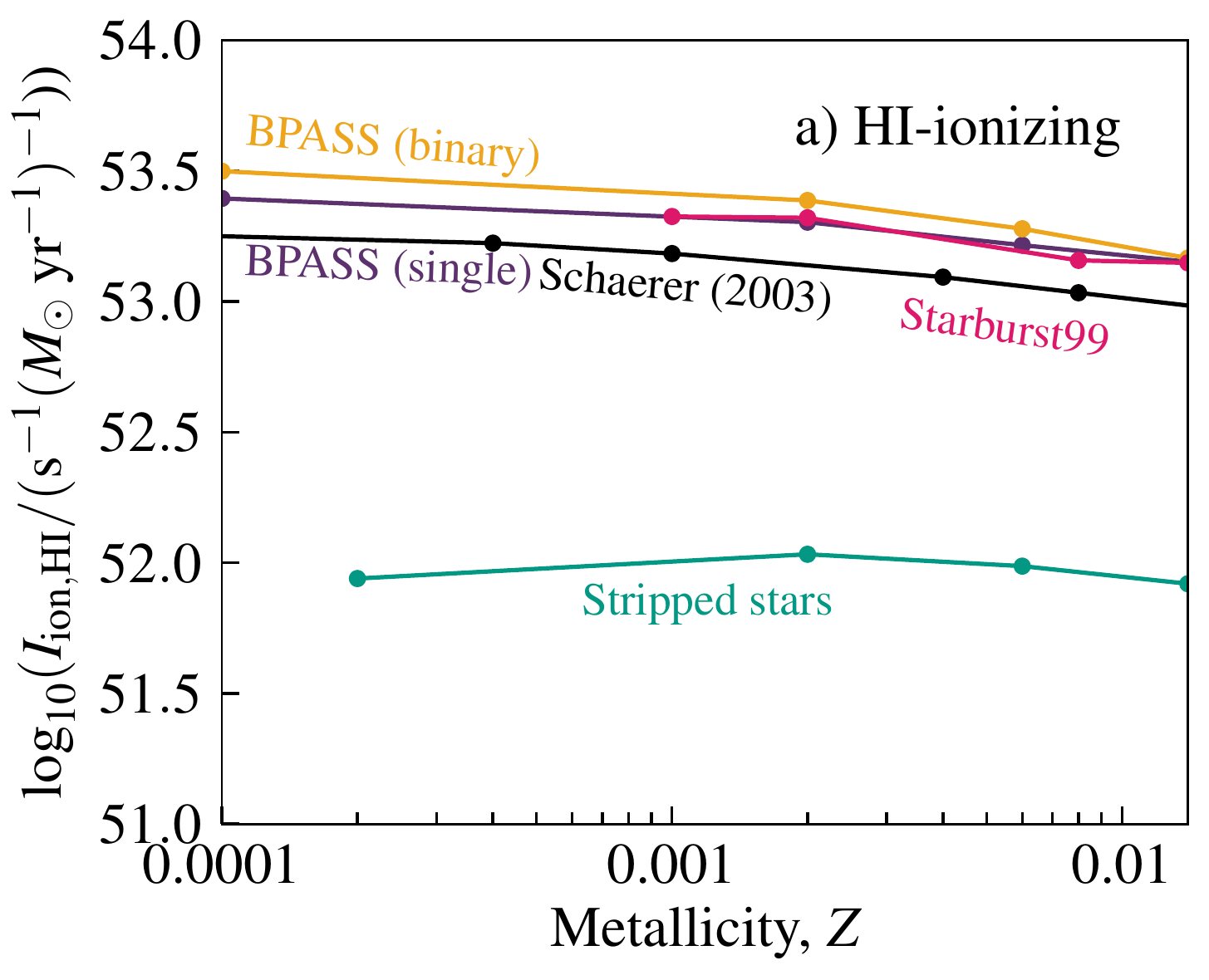}
\includegraphics[width=0.33\textwidth]{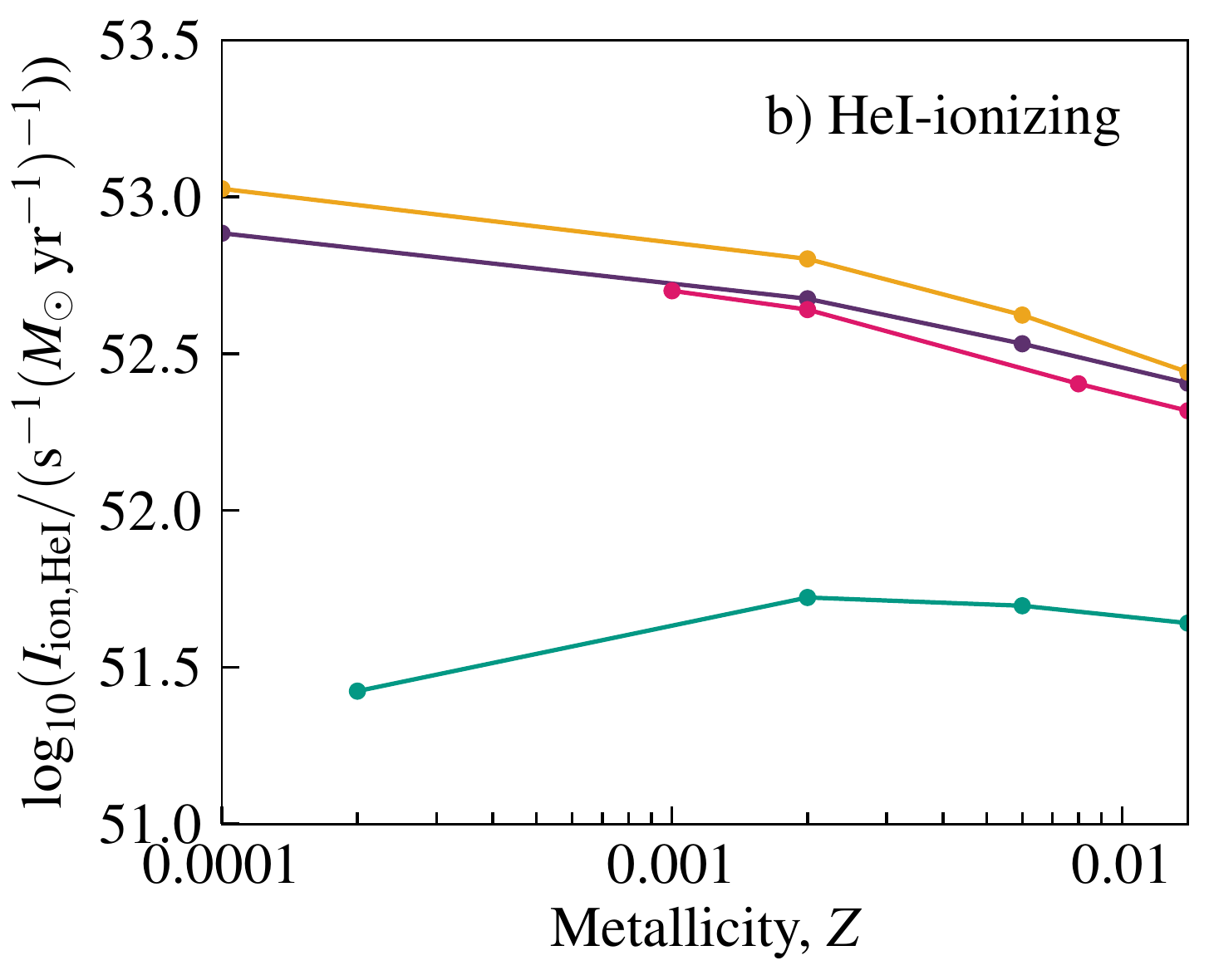}
\includegraphics[width=0.31\textwidth]{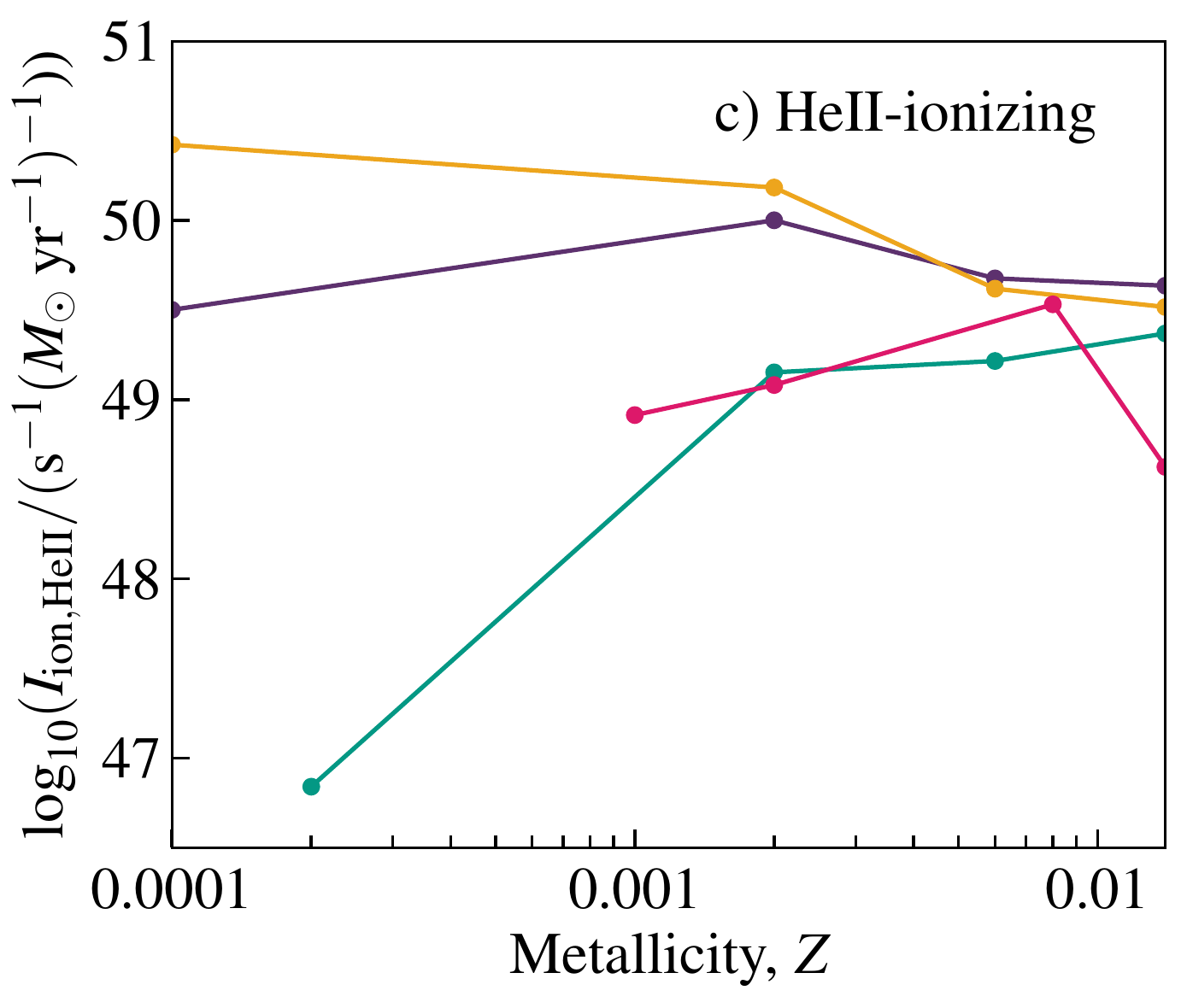}
\caption{The yields of \HI-, \HeI-, and \HeII-ionizing photons (\Iionz, \Iiono, and \Iiont respectively). We show these yields as functions of metallicity. The photon yields from stripped stars are shown in green and from massive stars in pink (using models from Starburst99). Panel a, b and c show the yield of \HI-, \HeI-, and \HeII-ionizing photons respectively. The values are presented in \tabref{tab:Iion}. ($Z_{\odot} \equiv 0.014$, \citealt{2009ARA&A..47..481A}.)}
\label{fig:Iion}
\end{figure*}

% Yield of ionizing photons is metallicity dependent
The yields of ionizing photons for stellar populations is dependent on metallicity. 
% yield = emission rate of ionizing photons for individual stars x their lifetime x number of stars formed + normalized by the total mass of stars formed
The reason for this dependence is multifaceted. The emission rate of ionizing photons increases with hotter and more luminous stars, and both the stellar temperature and the luminosity are dependent on metallicity. The stellar lifetime increases  with increasing metallicity, since the core masses are smaller at higher metallicity, causing the stars to burn slower. The  longer lifetime allows the star to radiate ionizing emission for a longer time.

% At lower metallicity the emission rate of ionizing photons first increases because the stars are more luminous, then decreases because they become cooler. Their lifetime decreases with increasing mass. The number of stripped stars is roughly similar at the different metallicities. Total effect is dominated by the emission rate of ionizing photons and therefore that shapes the yields of ionizing photons. 
In the case of stripped stars, the yields of ionizing photons are mostly determined by the emission rate of ionizing photons from the stars. The emission rate of \HI-ionizing photons from stripped stars increases with decreasing metallicity, reaching a peak at $Z\sim 0.002$. At even lower metallicity, our models predict that stripped stars are sufficiently cool for the emission rate of ionizing photons to be affected and to decrease again. This effect is subtle for \HI-ionizing emission. Stripped stars are the least efficient at producing \HI-ionizing photons at solar metallicity, with a yield of ionizing photons of $\Iionz \sim 7.8  \times 10^{51}$~s$^{-1}$  (\Msunyr)$^{-1}$. They are the most efficient at producing \HI-ionizing photons at $Z = 0.002$, reaching a 30\% higher value for the yield of \HI-ionizing photons.

% Present the figures and the table
This metallicity dependence on the yields of ionizing photons is presented in \tabref{tab:Iion} and \figref{fig:Iion}, where we display the yields of \HI-, \HeI-, and \HeII-ionizing photons at different metallicities. We show the yields for stripped stars using our models. For massive main-sequence and Wolf-Rayet stars, we show the predictions from the code \code{Starburst99} \citep{1999ApJS..123....3L, 2010ApJS..189..309L} and the single star version of the code \code{BPASS} \citep[version 2.2.1, \textit{Tuatara},][]{2017PASA...34...58E, 2018MNRAS.479...75S}. In \figref{fig:Iion}a, we also show the predictions from the models from \citet{2003AA...397..527S} since these are used for a similar calculation as ours that was presented in \citet{2014ARA&A..52..415M}. We also display the predictions from the version of \code{BPASS} that  accounts for interacting binaries. For all populations, we have assumed an initial mass function from \citet{2001MNRAS.322..231K} with the lower and upper mass limits of 0.1 and 100\Msun, respectively.

% Yield of HI-ionizing emission
\figref{fig:Iion}a shows that the yield of \HI-ionizing photons increases with decreasing metallicity for massive stars. This can be understood from the fact that stars are hotter and more luminous at lower metallicity, causing their emission rate of ionizing photons to increase. \tabref{tab:Iion} also shows that the predictions for \Iionz from the various codes agree relatively well. 

In the considered range of metallicities, the yields of \HI-ionizing photons from stripped stars is roughly 5\% of that from massive main sequence and Wolf-Rayet stars. Our models predict that their contribution is slightly larger at $Z = 0.002$. The models from \code{BPASS} predict that interacting binaries increase the yield of \HI-ionizing photons by about 5\% at solar metallicity. This is in agreement with what we predict stripped stars contribute with. At  $Z = 0.006$, \code{BPASS} predicts a larger difference of about 15\%, which is not in agreement with our models. The reason for this could be that \code{BPASS} includes stripped stars of higher masses than what we do. At $Z = 0.002$, \code{BPASS} predicts that binaries increase the yield of \HI-ionizing photons by $\sim 20\%$. At these low metallicities, \code{BPASS} assumes that rapidly rotating stars evolve chemically homogeneously, which causes an increase in \Iionz.

% Yields of HeI and HeII ionizing photons 
% HeI
The yield of \HeI-ionizing photons, \Iiono, follow a similar trend with metallicity as the yield of \HI-ionizing photons, as seen in \figref{fig:Iion}b. Down to $Z  =0.002$, the contribution from stripped stars to \Iiono is about 20\% compared to that from massive main sequence and WR stars. The reason why their contribution to \Iiono is relatively larger than to \Iionz is because stripped stars are hotter than main sequence stars. With the high temperatures of stripped stars, a larger fraction of their emitted \HI-ionizing photons are also \HeI-ionizing photons compared to massive main sequence and Wolf-Rayet stars. 

% HeII
\figref{fig:Iion}c shows that the yields of \HeII-ionizing photons, \Iiont, likely are uncertain for most stars. The predictions for massive main sequence and Wolf-Rayet stars from \code{Starburst99} and \code{BPASS} are different by up to an order of magnitude. The \HeII-ionizing emission is produced in the steep Wien-part of the stellar spectra, which is sensitive to temperature variations and the assumptions for wind mass loss. This can easily lead to large uncertainties in \Iiont. Taking the large uncertainties into account, we consider that stripped stars emit \HeII-ionizing photons at roughly the same or somewhat higher rate than massive main sequence and Wolf-Rayet stars. 

\end{document}